\documentclass[12pt]{iopart}

\usepackage{iopams}  
\usepackage{graphicx}
\begin{document}

\title[GW Generation via the Einstein-Langevin Equation]{Gravitational Wave Generation via the Einstein-Langevin Equation}

\author{Noah M. MacKay
\\ ORCID ID: 0000-0001-6625-2321}

\address{Institut für Physik und Astronomie, Universit\"at Potsdam,\\ Karl-Liebknecht-Str. 24/25, 14476 Potsdam, Germany}
\ead{noah.mackay@uni-potsdam.de}
\vspace{10pt}
\begin{indented}
\item[]\today
\end{indented}

\begin{abstract}
Detections of gravitational waves (GWs) since GW150914 has gained a contemporary interest in a potential quantum-classical correspondence between GWs and hypothetical gravitons. One such correspondence theory is stochastic gravity,  whereby graviton fluctuations are treated as the stochastic noise embedded in globally-flat manifolds and local gravitational interactions. Utilizing the Einstein-Langevin equation that describes graviton fluctuations, in attempt to form a correlation with GW generation, we utilize the hollow mass-shell model of coalescing compact binaries. This is to explore the second Newtonian postulate of neutralized internal gravitational fields, i.e. the stochastic noise of an enclosed, internal Minkowski manifold. This stochatic picture of GW formation implies the treatment of the enclosed gravitons as a Brownian bath. From the Einstein-Langevin equation, we establish a scaling relation where quanta dissipation depends inversely with the contracting volume (i.e., emission increases during coalescence). Using an Euler iteration scheme, we simulate the graviton fluctuations from inspiral to merger as a Wiener process, revealing a signal that qualitatively resembles macroscopic GW waveforms. While inherently heuristic and phenomenological, this approach provides a computational framework for exploring graviton-scale perturbations in GW formation. We discuss furthermore analytical waveform matching with the iteration scheme, as well as the justification of a Brownian analogy amidst current and state-of-the-art effective field theory frameworks. 
\end{abstract}

%
%
\submitto{\CQG}
%
\maketitle
%
%

\section{Introduction} \label{intro}

Since their discovery on September 14, 2015, gravitational waves (GWs) have been routinely observed by the LIGO-VIRGO-KAGRA (LVK) collaboration \cite{GWOSC, LIGOScientific:2018mvr, LIGOScientific:2021usb, KAGRA:2021vkt, LIGOScientific:2025slb}, now called the International Gravitational Wave Network (IGWN). Astrophysical sources of GWs have been well understood to be coalescing compact binaries (CCBs), which are behaviorally quadrupolar. Via the linearized Einstein field equations (EFEs) in a weak field limit, and modeling the CCB as two point-mass constituents, one obtains e.g. the circular traceless-transverse (TT) gauged spatial waveforms: 
\begin{equation}\label{gwsourced}
 h^\mathrm{TT}_{ij}=-\frac{4G\mu L^2\Omega^2}{D}\exp(i\,2\Omega t)\varepsilon^\mathrm{TT}_{ij},
\end{equation} 
where $D$ is the luminosity distance between the observer and the source, $\mu=m_1m_2/M$ is the reduced mass of the binary ($M=m_1+m_2$ is the total mass), $L$ is the binary separation, $\Omega$ is the orbital frequency,  $h_+\propto\cos(2\Omega t)$ and $h_\times\propto\sin(2\Omega t)$ via the $3\times3$ TT-gauged matrix $\varepsilon^\mathrm{TT}_{ij}$, and the strain amplitude is distinctly defined in source-dependent values. In this work, unless specified otherwise, $c=1$ is adopted. 

However, realistic CCBs complicate the simple waveforms by incorporating dynamic post-Newtonian (PN) \cite{Blanchet:2013haa} and perturbative post-Minkowskian (PM) \cite{Damour:2016gwp} corrections. Past the PN regime, CCBs undergo the inspiral-merger-ringdown (IMR) process, whereby the waveform intensifies with a dynamic frequency and amplitude enhancement, until reaching the maximum peak at the coalescence time $t_C$. Past coalescence, ignoring tidal deformations, the waveform dampens exponentially towards a zero flat-line. If one includes tidal deformations (e.g. for binary neutron stars (BNSs)), the 6PN correction in the orbital phase would influence the shape of the GW profile and the embedded dynamics to reach merger sooner than binary black holes (BBHs), with the waveform including post-merger, rapid-spin waveforms until reaching the BH plunge -- where the waveform flattens to zero.

Analytical approaches to model CCB dynamics and energetics include e.g. the state-of-the-art effective one-body (EOB) framework \cite{Buonanno:1998gg, Buonanno:2005xu}, in which coalescence is interpreted as the reduced mass $\mu$ spiraling down a spacetime well warped by the total mass $M$, and the more recent hollow mass-shell model \cite{MacKay:2024qxj, MacKay:2025uyg}. In the latter, the EOB ``marble-in-the-funnel" picture is reinterpreted as a contracting, rotating hollow shell whose exterior and time-dependent morphology is effectively Kerr-like. While the EOB formalism plays a central role in modern GW analysis -- albeit at the price of being computationally expensive --, the hollow mass-shell model offers a complementary, fully analytical description -- however at the expense of being an approximation. Despite its approximational nature, the mass-shell model for coalescence adopts a variational approach to the EFEs, combining a Laplace-Beltrami formulation of the Ricci tensor with a Kerr metric Ansatz. This construction yields a mass-shell surface energy that agrees well with the radiated energies inferred from GWTC observations \cite{MacKay:2025uyg}. Moreover, the hollow shell geometry naturally satisfies the Newtonian postulates for a spherical shell: (i) the external gravitational field of the shell is equivalent to that of a point mass, and (ii) the internel gravitational field vanishes. While Refs. \cite{MacKay:2024qxj, MacKay:2025uyg} primarily examined Postulate (i), the present work focuses on the implications of Postulate (ii). For comprehension, we review the CCB mass-shell model in Section \ref{model}.

\subsection{GWs in the Vacuum}

Regions in which the gravitational field vanishes are naturally identified as vacua. For GWs propagating in vacuum, we revisit the linearized EFEs, for which the energy-momentum tensor vanishes for no present source: $T_{\mu\nu}=0$, and the spacetime metric is a linearly-perturbed Minkowski metric: $g_{\mu\nu}=\eta_{\mu\nu}+h_{\mu\nu}$, with $|h|\ll1$. To linear order, we further imply $g^{\mu\nu}=\eta^{\mu\nu}-h^{\mu\nu}$ for the inverse metric to ensure $g^{\alpha\nu}g_{\beta\nu}=\delta^\alpha_{~\beta}$. Under these assumptions, the geometric half of the EFEs, $G_{\mu\nu}\equiv R_{\mu\nu}-Rg_{\mu\nu}/2=0$, yields an ungauged wave equation for the metric perturbation:
\numparts
\begin{equation}\label{waveq}
\frac{1}{2}\left(-\Box h_{\mu\nu}+\partial_{(\mu} V_{\nu)} \right)=0,
\end{equation}
where $\Box\equiv\partial_\alpha\partial^\alpha$ is the Minkowski d'Alembert wave operator,
\begin{equation} \label{4vec}
V_\mu:=\partial_\alpha h^\alpha_\mu-\frac{1}{2}\partial_\mu h^\alpha_\alpha,
\end{equation} 
\endnumparts
and $\partial_{(\mu}V_{\nu)}=\partial_\mu V_\nu+\partial_\nu V_\mu$ is a symmetric combination. At this stage, it is instructive to draw an analogy with electromagnetism (EM). An ungauged EM wave in the presence of a 4-current source satisfies $\Box A^\mu-\partial^\mu(\partial_\nu A^\nu)=\mu_0J^\mu$. In the vacuum, $J^\mu=0$, which reduces the ungauged wave equation to $\Box A^\mu-\partial^\mu(\partial_\nu A^\nu)=0$. The formal similarity with Eq. (\ref{waveq}) is evident, with the primary distinction arising from the tensorial rank of the metric perturbation. In EM, the Lorentz gauge condition $\partial_\nu A^\nu=0$ is naturally imposed, yielding the free wave equation $\Box A^\mu=0$ for photons \cite{griffiths}. In the gravitational case, the role analogous to $\partial_\mu A^\mu$ is played by the 4-vector $V_\mu$ via Eq. (\ref{4vec}). Imposing the Lorentz-like condition $V_\mu=0$ enforces invariance under infinitesimal coordinate transformations, and it defines the harmonic (de Donder) gauge \cite{Stewart:1991}. Under this condition, the vacuum GW equation reduces to the following, providing a plane-wave solution:
\begin{equation}\label{gwvacuum}
\Box h_{\mu\nu}=0\quad\Rightarrow\quad h_{\mu\nu}=A_0\exp(-ik_\alpha x^\alpha)\varepsilon_{\mu\nu}.
\end{equation}
Comparing Eqs. (\ref{gwsourced}) and (\ref{gwvacuum}), $A_0$ denotes the wave amplitude relating to the strain and source-dependent factors, $k_\mu=(\omega,\vec{k})$ is the 4-vector for the wave number, and $\varepsilon_{\mu\nu}$ is the polarization tensor.

Although $h_{\mu\nu}$ initially possesses ten independent components, relating to its degrees of freedom, the de Donder gauge removes redundant degrees of freedom, leaving two physically propagating modes. These correspond to the familiar TT-gauge polarizations, characterizing transversality, tracelessness, and the absence of temporal matrix components in $\varepsilon^\mathrm{TT}_{\mu\nu}$. In this sense, vacuum GWs behave qualitatively similar to vacuum EM waves, as photons also have two transverse polarization modes orthogonal to propagation. From a particle-physics perspective, vacuum GWs may therefore be interpreted as coherent-state gravitons (see e.g. Refs. \cite{Feynman:2002, Goldberger:2004jt, Goldberger:2006bd, Goldberger:2007hy, Kol:2007bc, Goldberger:2009qd, Foffa:2013qca, Rothstein:2014sra, Porto:2016pyg, Levi:2018nxp, Rafie-Zinedine:2018izq, Mogull:2020sak, Aoki:2024boe}), which are massless\footnote{If gravitons had a mass, it would be gauged to have an upper bound of $m_g\leq7.7\times10^{-23}\,\mathrm{eV}$ under the Compton wavelength $\lambda_g\geq1.6\times10^{13}$ km \cite{LIGOScientific:2017bnn}. However, massive gravitons would possess five physical degrees of freedom via $2s+1=5$ as spin-2 quanta, which contrasts from the TT-gauge essential to GWs. One may leverage ``ghost" gauge fields to possess any further excess degrees, but this would be unphysical to utilize. Therefore, gravitons are essentially massless \cite{Nakanishi:1979fg}.}, chargeless spin-2 quanta. Unlike photons, gravitons are expected to self-interact (a quality shared with strong-force-mediating gluons) -- see e.g. Ref. \cite{Rafie-Zinedine:2018izq, Mogull:2020sak} --, and it is precisely these non-linear interactions that give rise to perturbative corrections captured by PM expansion.

This quantum-classical correspondence (QCC) between GWs -- and more generally between the classical gravitational field -- and gravitons has been further explored by e.g. Cho and Hu \cite{Cho:2021gvg}, who interpret gravitons as quantum fluctuations in the relative separation between two masses. Returning to the CCB context, this perspective naturally motivates the intuition that gravitons are exchanged between the compact bodies as they inspiral and coalesce. This is consistent with effective field theoretical treatments of gravitational radiation (see e.g. Ref. \cite{Mogull:2020sak}). Within the CCB mass-shell construction introduced in Refs. \cite{MacKay:2024qxj, MacKay:2025uyg}, this intuition can be extended further. After distributing the compact bodies into an equatorial mass ring of reduced mass measure and integrating over all inclination angles to form a hollow shell, the resulting interior region may be viewed as a vacuum populated by graviton fluctuations. In this picture, the mass-shell interior is not empty but instead hosts a stochastic Brownian bath of self-interacting gravitons. A continued discussion can be found in Section \ref{shmear}.

\subsection{Motivating a Stochastic Brownian Picture}\label{sect:braun}

 The stochastic Brownian framework for vacuum GWs, interpreted as fluctuations of an underlying graviton field, is motivated by early work on stochastic gravity, notably Ref. \cite{Moffat:1996fu}. In this approach, quantum fluctuations of an otherwise flat Minkowski spacetime induce small, stochastic variations in the gravitational field, which appear as proportionally small corrections to the classical EFEs. When such fluctuations are incorporated consistently, the resulting dynamics naturally take the form of Langevin \cite{Hu:1994ep} and Fokker-Planck \cite{Moffat:1996fu} equations, describing first-order metric perturbations as graviton fluctuations under a QCC rooted in wave-particle (as well as field-particle) duality. 

Adopting this stochastic viewpoint, the CCB mass-shell -- whose enclosed volume $V(t)$ decreases as the system approaches coalescence -- allows for a heuristic interpretation of the contained gravitational bulk as an increasingly energetic, ultra-relativistic graviton gas. This picture is consistent with the well-known, time-dependent enhancement of GW frequency and amplitude across the inspiral and merger phases. Within this framework, individual gravitons are treated statistically, with their random motion governed by stochastic dynamics. Importantly, the stochasticity of the graviton bath is not attributed to the ambient cosmological background temperature of order $T_\mathrm{CMB}\sim1\mathrm{K}$, but rather to fluctuations sourced by the inspraling compact bodies themselves, as well as high-energy graviton-graviton interactions that become increasingly relevant near merger (see e.g. \cite{Rafie-Zinedine:2018izq, DeWitt:1967uc,  Blas:2020och, Delgado:2022uzu, Herrero-Valea:2022lfd} on graviton-graviton scattering amplitudes in the general sense). 

Such confined graviton fluctuations may be described by the Einstein-Langevin equation \cite{Hu:1994ep}: an integro-differential extension of the EFEs that incorporates dissipative and stochastic effects. Although originally formulated for a gobally-flat, cosmological setting -- where the dissipation kernel depends on the Friedmann-Robertson-Walker (FRW) scalar factor and the Hubble parameter -- the underlying formalism may be applied to locally-flat systems. More specifically in this work, the Einstein-Langevin equation is employed as a stochastic description of graviton fluctuations in the CCB mass-shell interior, where the shell volume contracts progressively until merger. From this perspective, quantities analogous to the cosmological Hubble parameter serve to encode dissipation arising from more local volumetric fluctuations, i.e. those induced by mass-shell contraction during coalescence. 

By contrast, classical Brownian motion is governed by the Langevin equation \cite{lang}, which balances deterministic forces sourced by a potential gradient $-\vec\nabla U(x)$, velocity-dependent damping $\gamma \vec{v}$, and Gaussian noise $\sigma\zeta_2(t)$. The formal similarity between the Langevin equation and the Einstein-Langevin equation, besides the homage to Paul Langevin, provides a natural bridge for modeling graviton-scale fluctuations using stochastic methods. Treating the gravitons confined within $V(t)$ as a Brownian bath offers, therefore, a framework in which the emergence of GWs -- from inspiral to merger -- can be mapped onto stochastic graviton dynamics. Exploring this correspondence is the central aim of this work. 

\subsection{Paper Structuring}
 
 While dynamical GWs can be described analytically and numerically, this study adopts an iterative approach to model graviton fluctuations across coalescence as a Wiener process, from inspiral to merger, where GW emission peaks. To implement this computationally, a discretization scheme akin to those commonly used in Langevin-equation-based modeling \cite{Yuvan2021, Yuvan2022ent, Yuvan2022sym, Bier2024, MacKay:2024} is employed. Numerical iteration, however, depends critically on the effective potential energy profile of the Brownian bath, which determines the dissipation force $-\vec{\nabla}U(x)$. For gravitons governed by the Einstein-Langevin equation, this potential is derived from the quanta dissipation kernel; its evaluation is presented in Section \ref{anres}.
  
A secondary objective of this study is to examine the role of Gaussian noise in a discretized Einstein-Langevin framework during binary coalescence. Near peak GW emission, and under the assumption of effective thermal equilibrium between the stochastic, internal system and the external background, the iterated fluctuations within the contracting volume serve as a proxy for stochastic fluctuations of the GWs emitted from the shell surface. The numerical results may indicate that, within the stated assumptions, graviton-scale fluctuations can produce signal patterns that qualitatively resemble gravitational waveforms. Therefore, this framework is intended as a proof-of-concept rather than a precise model of predicting observable GWs. For reproducibility, the \textsc{Wolfram Mathematica} implementation of the Euler iteration scheme is presented in Section \ref{numres}. An extended discussion follows in Section \ref{disc}, where we review select analytical consequences to the iterative/Brownian approach, and we conclude in Section \ref{concl}.

\section{Methods}

\subsection{Reviewing the CCB Mass-Shell Model} \label{model}

For both stable classical binaries and CCBs, the reduced mass $\mu$ simplifies two-body dynamics into a singular effective system. For CCBs specifically, the binary may be assumed to behave as a singular object (i.e. a rotating and contracting hollow mass shell) when viewed from far away. While undergoing coalescence, the rate of change in the separation $-d{L}/dt$ w.r.t the observer time relates to the mass shell's contracting  diameter with the rate of change $-d{S}/dt$ w.r.t the observer time. When integrated over the timelapse $t'\in[t,\,t_C]$, where $t$ is dynamic and $t_C$ is a fixed coalescence time, we yield:
\begin{equation}\label{seps}
-S(t_C)+S(t)=-L(t_C)+L(t).
\end{equation}
In Eq. (\ref{seps}), the CCB separation and shell diameter at the coalescence time $t_C$ are fixed values. Adopting EOB convention by imposing coalescence ends when the masses make contact: $L(t_C)=r_1+r_2$, under the shell diameter that identifies the total mass horizon diameter: $S(t_C)=4GM$, we define:
\begin{equation}\label{rads}
S(t)=L(t)-r_1-r_2+4GM.
\end{equation}
In this model, CCBs are viewed instead as a shrinking mass shell with the constant measure $\mu$ and contracting radius $\rho(t)=S(t)/2$, until reaching the ``innermost'' shell that is the total mass horizon. Thus, the volume of this mass shell is $V(t)=4\pi \rho(t)^3/3$.

\subsubsection{Mass-Shell Waveforms}

Even in the CCB mass-shell model, the respective TT-gauged spatial waveform of this mass shell model is conventionally quadrupolar:
\begin{equation}\label{wave}
h^\mathrm{TT}_{ij}=\frac{2G}{D}\ddot{Q}_{ij}^\mathrm{TT},
\end{equation}
where $Q_{ij}$ is the quadrupole moment tensor. For a CCB treated as an effective compact object rotating at the dynamic orbital frequency $\Omega=\Omega(t)$,  the quadrupole moment is proportional to the point mass moment of inertia $I=\mu \rho^2$. If we assume e.g. (nearly-)circular orbits, and demonstrate time variation in the CCB separation $S=S(t)$ and $\Omega$ while conserving angular momentum $J=I\Omega$, we eventually yield waveform expressions written entirely in terms of observer-time derivatives of e.g. $\Omega$, c.f. Refs. \cite{MacKay:2024qxj, MacKay:2025uyg}:
\numparts
 \begin{eqnarray}\label{hp2}
&h_{+}^{\mathrm{TT}}=-\frac{G\mu S^2}{D}\Big(\sin(2\Omega t)\left(-\frac{\dot{\Omega}}{\Omega}\left(\Omega +t\dot{\Omega}\right)+\dot{\Omega}+\frac{t}{2}\ddot{\Omega}\right)\\\nonumber
&\quad\quad\quad\quad\quad\quad\quad -\cos(2\Omega t)\left(-\frac{\ddot{\Omega}}{4\Omega}+\frac{\dot{\Omega}^2}{4\Omega^2}-\left(\Omega+t\dot{\Omega}\right)^2\right)-\frac{\ddot{\Omega}}{4\Omega}+\frac{\dot{\Omega}^2}{4\Omega^2}\Big),\\\label{hc2}
&h_{\times}^\mathrm{TT}=-\frac{G\mu S^2}{D}\Big(\sin(2\Omega t)\left(-\frac{\ddot{\Omega}}{4\Omega}+\frac{\dot{\Omega}^2}{4\Omega^2}-\left(\Omega+t\dot{\Omega}\right)^2\right)\\\nonumber
&\quad\quad\quad\quad\quad\quad\quad +\cos(2\Omega t)\left(-\frac{\dot{\Omega}}{\Omega}\left(\Omega +t\dot{\Omega}\right)+\dot{\Omega} +\frac{t}{2}\ddot{\Omega}\right)\Big).
\end{eqnarray}
\endnumparts

Given that the orbital expressions are directly coupled to the wave profiles, and thus govern the evolution of the waveform envelope during coalescence, one must also consider any post-merger dynamics. For times $t>t_C$, the GW signal transitions into its ringdown phase, which may be modeled using quasi-normal modes and piecewise stiching, e.g. with a Gaussian-like damping inspired by stellar collapse and post-merger relaxation \cite{Buonanno:2005xu, Gundlach:1993tp}. In the general sense, taking Eqs. (\ref{hp2}) and (\ref{hc2}) as the inspiral-merger basis, the complete IMR waveform may be expressed as a linear expansion in sine and cosine profiles, each scaled by a time-dependent, polarization-specific envelope function $\mathfrak{E}(t)$ encapsulating the dynamics:
\begin{equation} \label{pieces}
h_{+/\times}^\mathrm{TT}(t)=-\frac{G\mu S^2}{D}\left[\cos(2\Omega t) \mathfrak{E}_{+/\times}^\mathrm{cos}(t)+\sin(2\Omega t)\mathfrak{E}_{+/\times}^\mathrm{sin}(t)\right].
\end{equation}
E.g., for the simple wave approximation whereby $\dot{\Omega},~\ddot{\Omega}\rightarrow0$, $\mathfrak{E}(t)$ reduces uniquely into a constant (either $\Omega^2$ or 0, depending on polarization) for all $t$, and the waveform is scaled only by its amplitude, similar to Eq. (\ref{gwsourced}). In contrast, the fully dynamic IMR amplitude is encoded in a characteristic piecewise structure of $\mathfrak{E}(t)$; for times $t\leq t_C$, this is  given by the complete expressions in Eqs. (\ref{hp2}) and (\ref{hc2}). For times $t>t_C$, this transitions into a Gaussian-like damping form, motivated by Ref. \cite{Gundlach:1993tp}, resulting to the asymptotic level-out to zero (see footnote\footnote{For BNSs in particular, the envelope function $\mathfrak{E}(t)$ acquires additional tidal contributions entering through the PN corrections to $\Omega$ up to merger, followed by non-trivial post-merger scaling prior to a possible BH collapse.}). 

\subsubsection{Mass-Shell Surface Energy}\label{shelen}

While it is conventional to extract the energy density (and energy flux density) of GWs via $h_+^2+h_\times^2$, using either both Eqs. (\ref{hp2}) and (\ref{hc2}) or just Eq. (\ref{pieces}) generally, a more heuristic approach was employed in Refs. \cite{MacKay:2024qxj, MacKay:2025uyg}, wherein the relevant quantity was obtained through a variational treatment of the EFEs. Per the EFEs $G_{\mu\nu}=8\pi GT_{\mu\nu}$, the energy density of the CCB mass shell is given as follows:
\begin{equation}\label{efes}
T_{00}=\frac{1}{8\pi G}G_{00},
\end{equation}
where $G_{00}$ encodes the geometric response via the Ricci component $R_{00}$, the metric component $g_{00}$, the Ricci scalar $R=g^{\mu\nu}R_{\mu\nu}$, relevant Christoffel symbols $\Gamma^\alpha_{\mu\nu}$ and the full metric tensor. The additive cosmological contribution $\Lambda g_{\mu\nu}$ is neglected here. For a CCB modelled as a rotating and contracting mass-shell, it would be cumbersome to solve the EFEs in the conventional, hierarchial sense (i.e., beginning with a known source and solve for the metric). Because it is the objective to find $T_{00}$, we instead begin with a well-defined geometric Ansatz for the spacetime configuration. This unusual, inverted procedure is analogous to quantum mechanical variational method, where one solves an impossible Hamiltonian by first assuming an Ansatz wavefunction, then evaluating the expectation value of the Hamiltonian to obtain an energy functional, and lastly minimizing the functional to obtain the anticipated ground state energy. 

In the context of CCB as a mass shell, we follow a similar logic by first assuming a well-defined metric Ansatz that best describes the system -- one that is already a known solution to the EFEs --, on which we apply differential operators inherent in the Christoffel symbols $\Gamma\sim g^{-1}\partial g$ and the Ricci tensor $R\sim\partial\Gamma+\Gamma\Gamma$ (suppressing the indices), and define $G_{\mu\nu}\propto T_{\mu\nu}$ component-wise, focusing exclusively on $G_{00}\propto T_{00}$. Through these schematic structures for $\Gamma^\alpha_{\mu\nu}$ and $R_{\mu\nu}$, we claim that the Ricci tensor can be effectively defined as a Laplace-like operator acting on the metric tensor: $R\sim \nabla(g^{-1}\partial g)\sim\nabla^2g$, where $\nabla\sim(\partial+\Gamma)$ is the covariant derivative (see footnote\footnote{One might be wary of this definition, given metric compatibility $\nabla_\mu g^{\mu\nu}=0$. However, it is shown that the covariant Laplacian: the Laplace-Beltrami operator, is coordinate-dependent and provides non-trivial results when acted on specific metric tensor components.}). This interpretation is consistent with Lemma 3.32 from Chow and Knopf \cite{Chow:2004}, which states in leading order:
\begin{equation}\label{ricci}
R_{\mu\nu}\simeq-\frac{1}{2}\Delta^\mathrm{LB} g_{\mu\nu}\,\left(+~\mathrm{lower~order~terms}\right).
\end{equation}
The operator $\Delta^\mathrm{LB}$ is the Laplace-Beltrami operator, which for a well-defined metric with a non-zero determinant may be written in the Christoffel symbol-free form as:
\begin{equation}\label{lb}
\Delta^\mathrm{LB}=\frac{1}{\sqrt{-g}}\,\partial_\alpha\left(\sqrt{-g}g^{\alpha\beta}\partial_\beta \right),
\end{equation}
where $\sqrt{-g}=\sqrt{-\mathrm{det}(g_{\mu\nu})}$ and $g^{\mu\nu}$ is the inverse metric. From Eq. (\ref{efes}), it follows that $G_{00}$ depends on $R_{00}$, $R$, and $g_{00}$, and through Eq. (\ref{ricci}) $R_{00}\propto g_{00}$. In this Laplace-Beltrami formulation, the effective energy density is defined approximately however essentially through $g_{00}$, dropping the lower order terms in Eq. (\ref{ricci}):
\begin{equation}\label{endens}
T_{00}\approx -\frac{1}{8\pi G}\left(\frac{1}{2}\Delta^\mathrm{LB} +\frac{1}{2}R \right)g_{00}.
\end{equation}
One can see that, as $g_{00}$ is negative under the $(-,+,+,+)$ metric signature, this approximation yields a positive calculation for $T_{00}$. 

As detailed in Refs. \cite{MacKay:2024qxj, MacKay:2025uyg}, the Laplace-Beltrami treatment to the EFEs depends on the choice of metric Ansatz and how one approaches the Ricci scalar as a result. For a CCB mass shell assumed to resemble a spinning, compact object, our choice of Ansatz metric is the Kerr metric \cite{Kerr:1963ud, Boyer:1967, Chandrasekhar:1985kt} in Boyer-Lindquist coordinates, with the mass measure being the reduced mass of the CCB, and the Ricci scalar instead being effectively defined by the Kerr-metric Kretschmann scalar: $K=R_{\alpha\beta\mu\nu}R^{\alpha\beta\mu\nu}$ \cite{dInverno:1992gxs, Henry:1999rm, Visser:2007fj}. After one finds the complete, effective expression for $G_{00}$, we Taylor expand the expression under a small Kerr spin-parameter $a$ (recalling that the final spin parameter of CCBs is typically $a<1$ \cite{LIGOScientific:2018mvr, LIGOScientific:2021usb, KAGRA:2021vkt, LIGOScientific:2025slb}, see footnote\footnote{In the GWTC catalogs, the dimensionless spin parameter is defined as $\chi\equiv a/(GM)$. Since $a=\rho\beta$ on the mass-shell surface, where $\beta$ is the rotational velocity ratio across coalescence, we obtain $a_C=\rho_C\beta_C\lesssim4GM/5$ at $t=t_C$. Therefore, at merger, $\chi_C=a_C/(GM)\lesssim4/5$, which is consistent with the observed clustering of final dimensionless spins in the range $\chi\sim0.6-0.8$ across GWTC events. This indeed satisfies the Kerr bound $\chi<1$.}) and integrate over the polar angles $\theta\in[0,\pi]$ (an analytical consequence of the model's shell geometry, whereby the inclination angle $\iota\in[0,\pi]$ serves as the effective polar angle). This yields, via Eq. (\ref{endens}):
\begin{equation} \label{endens1}
T_{00}\simeq\frac{1}{4}\frac{G\mu^2}{r^4}\left(1.577-8.320 \frac{a^2}{r^2}\right),
\end{equation}
and we obtain the energy $E=T_{00}V$ at the shell radius $r=\rho$. We further define the surface energy at the time of merger $t=t_C$, for which $\rho=2GM$ and $a/\rho=\beta_C\propto GMf_\mathrm{GW,peak}$:
\begin{equation}\label{energy}
\Rightarrow\quad E(t_C)\simeq 0.826\frac{\mu^2}{M}\left(1-5.276 \beta_C^2 \right).
\end{equation}
Furthermore, one may determine the maximum coalescence rotational velocity permitted by this approximation, in order to avoid unphysically negative energy values. The energy given by Eq. (\ref{energy}) vanishes at $\beta_C^2\simeq0.190$, which sets an upper bound, ceiling value on the normalized velocity of $v_C\leq0.435 c$. This is comparable to the rotational velocity at the total mass ISCO radius: $v_\mathrm{ISCO}=0.408c$, which may serve as an extra constraint.

\subsection{The Einstein-Langevin Equation} \label{eldisc}

The Einstein-Langevin equation describes first-order metric perturbations on a flat background as graviton fluctuations confined within a closed volume $V$. Using FRW variables, it is written as
\begin{eqnarray}\label{eleq}\nonumber
\ddot{a}&-\frac{2}{3}\Lambda a^3\\
&+\frac{\hbar G}{12\pi a}\int_{\tau_0}^{\tau}d\tau'H(\tau')\int_0^\infty dk\,k^3\cos\left[k(\tau-\tau')\right]=\frac{4\pi G\hbar}{3V a}\dot{\zeta}_2(\tau),
\end{eqnarray}
where $a=a(\tau)$ and $\tau=\int dt/a$ is the conformal time (in other literature, e.g. Ref. \cite{Hu:1994ep}, the conformal time is denoted as $\eta$). Also, $H=\dot{a}/a$ is the Hubble parameter with $^\bullet \equiv d/d\tau$, and the cosmological constant $\Lambda$ is present.

The double integral defines the quanta dissipation kernel. The integral over $\tau'$ runs from the initial-reference conformal time $\tau_0$ to the current, dynamic conformal time $\tau$. This kernel encodes both the spectral contributions of perturbations (via the $k^3$-weighting) and their conformal-time-dependent dissipation dynamics. On the right-hand side, the Gaussian noise generator $\zeta_2(\tau)$ is acted upon by a conformal time derivative. Its amplitude is shown to be proportional to the canonical quantum of area $A\propto8\pi\hbar G$, which emerges from independent loop quantum gravity frameworks \cite{Rovelli:1987df, Rovelli:1989za, Rovelli:1994ge, Rovelli:2004}. This proportionality provides a natural connection between stochastic fluctuations and the discretized structure of spacetime itself.

To investigate the role of stochastic fluctuations in GW generation, we model graviton fluctuations as a Brownian system confined within the rotating, contracting volume $V=V(t)$. This analogy treats graviton fluctuations as stochastic perturbations induced by high-energy self-interactions and binary coalescence, rather than by thermodynamic means, as discussed in Section \ref{sect:braun}. For numerical analysis, we reduce Eq. (\ref{eleq}) to a first-order differential form, suitable for discretization via a forward Euler scheme. In doing so, we first neglect the cosmological term containing $\Lambda$, see footnote\footnote{This is consistent with the external analysis of the CCB mass-shell, which also ignores $\Lambda$, thereby constraining the system to a local setting.}. Then, we evaluate both sides over $d\tau$ to obtain:
\begin{eqnarray} \label{eleq2}
\dot{a}+\frac{\hbar G}{12\pi}\int_0^\infty \frac{d\tau}{a(\tau)} \int_{\tau_0}^{\tau}d\tau'H(\tau')&\int_0^\infty dk\,k^3\cos\left[k(\tau-\tau')\right]\nonumber\\
&=\frac{4\pi \hbar G}{3Va}\zeta_2(\tau).
\end{eqnarray}
After integration over $d\tau$, the conformal-time differentiation on $\zeta_2(\tau)$ is removed, restoring a standard Gaussian noise term. Additionally, the dissipation kernel evaluated at $\tau$ integrates out the time dependence and yields a constant force factor $F$. Consequently, Eq. (\ref{eleq2}) takes a form analogous to a Langevin equation with constant dissipation:
\begin{equation}\label{langeq}
\gamma\frac{d}{dt}{x}(t)+F\Theta(x)=\sigma{\zeta}_2(t),
\end{equation}
where $x=x(t)$ represents the fluctuation displacement and the Heaviside function $\Theta(x)$ ensures that dissipation remains positive. The corresponding potential energy profile for these fluctuations is a barrier-linear well, defined as
\begin{equation} \label{barlinpot}
U(x)=F\cdot
\left\{
\begin{array}{ll}
\infty & \mathrm{for }\quad x \leq 0 \\
x      & \mathrm{for }\quad x > 0
\end{array}
\right. .
\end{equation}

To simulate graviton fluctuations numerically, we discretize Eq. (\ref{eleq2}) using a forward Euler scheme, approximating conformal time in discrete steps: $d\tau\rightarrow\Delta \tau$. This method captures the contributions of both the dissipation force -- originating from the three-fold integral kernel -- and the Gaussian noise $\zeta_2(\tau)$ at each timestep, effectively representing the graviton's stochastic motion across the evolving system as a Wiener process.

By treating the fluctuating graviton displacement as a Wiener process, the numerical approach remains computationally stable and efficient even over a large numbers of timesteps. Conceptually, this simulates the random walk of a single representative graviton, with each ``kick" corresponding to interactions or collisions with neighboring gravitons. If we assume that gravitons propagate in a coherent state -- as implied for GWs -- the one-particle mapping can be amplified by the occupation number of the coherent state. This scaling enables the Wiener process of a single graviton to serve as a proxy for the collective, macroscopic GW signal. Consequently, the method provides a qualitative proof-of-concept for how individual graviton fluctuations may contribute to the coherent, observable GW signal.

\subsection{Discretization and Numerics} \label{discrete}

To simulate the random walk of a particle governed by e.g. Eq. (\ref{langeq}), we model the dynamics as a series of discrete ``kicks'' (or timesteps) within the potential well $U({x})$ \cite{MacKay:2024}. Each kick is introduced by the Gaussian noise generator $\zeta_2(t)$, while the negative gradient of the potential, $-\vec\nabla U(x)$, drives the particle back toward equilibrium at $x=0$. The stochasticity of these kicks follows a Gaussian distribution, forming a diffusive sequence of timesteps that represents the particle jittering at each moment. For numerical simulations, Eq. (\ref{langeq}) is discretized by approximating derivatives over finite intervals:
\begin{equation}
\Delta {x}_i=\frac{\Delta t}{\gamma}\left[-{F}\Theta({x}_i)+\sigma\zeta_2(t_i)\right],
\end{equation}
where $\Delta {x}_i={x}_{i+1}-{x}_i$. Here, the Gaussian noise at timestep $i$ is sampled as $\zeta_2(t_i)=\theta_{i,2}(\Delta t)^{-1/2}$, with $\theta_{i,2}$ drawn from a normalized Gaussian distribution. This yields the standard forward Euler iteration scheme \cite{Yuvan2021, Yuvan2022ent, Yuvan2022sym, Bier2024}:
\begin{equation} \label{euler}
{x}_{i+1}={x}_i+\frac{\Delta t}{\gamma}\left[-{F}\Theta({x}_i)+\sigma\frac{\theta_{i,2}}{\sqrt{\Delta t}} \right].
\end{equation} 

In simple cases, the damping factor $\gamma$ is often scaled to unity, highlighting the undamped noise-driven dynamics. Simulation parameters include the number of timesteps $l$, which also serves as the length of the simulation, the timestep size and $\Delta t$, and the initial position $x_0$ (e.g., ${x}_0=10^{-3}$) to avoid numerical infinities at $x=0$. 

It is important to note that Eq. (\ref{euler}) can become inaccurate if the force $F$ is large relative to the noise term. A strong force can anchor the particle near $x_0$ and suppress Brownian jitter. To maintain accuracy without artificially reducing $F$, the timestep size $\Delta t$ should be sufficiently small so that $\Delta x_i=x_{i+1}-x_i$ remains proportionally small. This ensures the discrete iterations closely approximate the continuous dynamics of Eq. (\ref{langeq}). The trade-off is that a smaller $\Delta t$ requires a larger number of timesteps $l$ to simulate a sufficiently long realization of the stochastic process (e.g., $l=10^5$ for $\Delta t=10^{-3}$).

\section{Analytical Results} \label{anres}

\subsection{The Quanta Dissipation Kernel} \label{quantder}

In Eq. (\ref{eleq2}), the three-fold integral defining the quanta dissipation kernel, denoted here as $\mathcal{K}_3$, is given by
\begin{equation} \label{3kern}
\mathcal{K}_3=\int_0^\infty \frac{d\tau}{a(\tau)} \int_{\tau_0}^{\tau}d\tau'H(\tau')\int_0^\infty dk\,k^3\cos\left[k(\tau-\tau')\right].
\end{equation}
As discussed in Section \ref{eldisc}, integration over $d\tau$ requires that the kernel ultimately yields a conformal-time-independent scontribution. This requirement ensures that the effective potential governing the stochastic flucutations retains the barrier-linear form described as Eq. (\ref{barlinpot}), allowing $\mathcal{K}_3$ to be identified with a constant force scale $F$. Evaluating the kernel furthermore assumes an up-close observer perspective, whereby one can ``observe" how the CCB mass-shell morphology impacts the behavior of the dissipation kernel. 

\subsubsection{Integral with respect to $k$}

The innermost integral over $k$ corresponds to the Fourier cosine transform of the function $k^3$, which is generally defined for any function $f(k)$ as \cite{wolfram}:
\begin{equation}
\mathcal{F}_{\mathrm{cos}} [f(k)]=\sqrt{\frac{2}{\pi}}\int_{0}^\infty dk\,f(k)\cos\left[k\xi\right].
\end{equation}
For power-law functions $f(k)=k^n$ with whole integer $n$, and provided that $\xi=\tau-\tau'$ for our case of evaluating the innermost integral, the transform depends on the oddity of $n$:
\begin{equation}
\mathcal{F}_{\mathrm{cos}}[k^n]=\left\{
\begin{array}{ll}
  i^n\sqrt{2\pi}\delta^{(n)}(\tau-\tau'), & \mathrm{if}~n~\mathrm{even,}\\
  n!\sqrt{\frac{2}{\pi}}\cos\left[\frac{\pi}{2}(n+1)\right](\tau-\tau')^{-(n+1)}, &
  \mathrm{if}~n~\mathrm{odd.} 
     \end{array}\right.
\end{equation}
In the present case, $n=3$. Accounting for the normalization factor $\sqrt{\pi/2}$, the $k$-integration reduces Eq. (\ref{3kern}) to
\begin{equation} \label{2kern}
\mathcal{K}_3=6\int_0^\infty \frac{d\tau}{a(\tau)}\int_{\tau_0}^{\tau}d\tau'\frac{H(\tau')}{(\tau-\tau')^{4}}.
\end{equation}

\subsubsection{The Hubble Parameter for Binary Coalescence} \label{hubbles}

The integral over $\tau'$ becomes ambiguous if the Hubble parameter $H(\tau')$ is left unspecified. For our given system of a CCB mass-shell model, $H(\tau')$ is therefore defined directly from the dynamics involved in mass-shell volumetric contraction, as discussed in Section \ref{sect:braun}. 

Using the conventional Hubble relation $\vec{u}=H\vec{r}$ \cite{Hubble:1929ig}, the radial velocity $\vec{u}$ of the contracting volume is modeled after the inward-pointing radial velocity of the CCB itself. Throughout coalescence, this velocity may be expressed as $\vec{u}=-2\beta^5(GM/P)^{1/2}\hat{r}$ (c.f. Ref. \cite{Loutrel:2018ssg}), where $\beta=|\vec{v}|/c\leq0.435$ is the normalized rotational speed ratio introduced earlier in Section \ref{shelen}. Here, $\beta^5$ serves as an effective parameterization of the osculating eccentricity, reflecting the tendency for eccentricity to increase dynamically as coalescence proceeds. Correspondingly, the semi-latus rectum $P$ is roughly $6GM$ for nearly circular orbits and $(10\sim15)GM$ for orbits with higher eccentricity. The radius vector $\vec{r}$, representing the mass shell's scale, has the magnitude proportional to $V^{1/3}$ and therefore can be modeled as $\rho(t)=S(t)/2$ via Eq. (\ref{rads}).

As discussed in Section \ref{model}, both contraction and rotation of the CCB mass-shell evolve over the observer time $t$ throughout coalescence. Consequently, the radial velocity $\vec{u}\propto\beta^5 P^{-1/2}$ explicitly becomes a function of $t$ through the evolution of $\beta(t)$ and the dimensionless semi-latus rectum $\widetilde{P}(t)=P(t)/(GM)$. We therefore define an effective, observer-time-dependent Hubble parameter for the CCB mass-shell interior:
\begin{equation} \label{hubble}
H(t)\equiv\frac{\vec{u}(t)}{\vec{r}(t)}=-\frac{2\beta(t)^5}{\rho(t)\sqrt{\widetilde{P}(t)}}.
\end{equation}
This quantity is negative, reflecting the contraction of the enclosed volume $V$ during coalescence. The reciprocal of its absolute value, $\Delta t=1/|H(t)|$, provides an order-of-magnitude estimate for the remaining coalescence timescale from a given observational start time $t_0$, which can be inferred by starting conditions of the observable parameters. 

E.g., for equal binary masses on the order $m_1=m_2\sim10^{31}$ kg (i.e., $M\sim2\times10^{31}$ kg), suppose we begin observation at the start of inspiral where $\beta\simeq0.01$ and $\widetilde{P}\approx6$. The time to merger at the start of inspiral, at leading Newtonian order, is also given as proportional to the fourth power of the seperation distance $L$, at which moment $L\approx S$ in the mass-shell interpretation for large separations:
\begin{equation} 
\Delta t_{\mathrm{ins}}\equiv (t_C-t_0)=\frac{5}{256}\frac{S^4}{G^3m_1m_2M}. 
\end{equation}
Using the above to solve for $\rho=S/2$ and substituting it in Eq. (\ref{hubble}) while calling $1/|H|=\Delta t_\mathrm{ins}$, we yield an inspiral timelapse expression:
\begin{equation}\label{ins}
\Delta t_\mathrm{ins}^{3/4}=\left(\frac{16}{5} \right)^{1/4}\frac{\sqrt{\widetilde{P}}}{2\beta^5}\left(G^3m_1m_2M \right)^{1/4},
\end{equation}
where $\Delta t_\mathrm{ins}$ is isolated by imposing the $4/3$ power on both sides of Eq. (\ref{ins}). For our values for $\beta\simeq0.1$ and $\widetilde{P}\approx6$ and our binary mass scalings, we yield a timelapse until merger of approximately 278.965 seconds, or roughly 4 minutes and 39 seconds. This characteristic timescale is comparable to that calculated by another relation dependent on the GW frequency and the binary's chirp mass $\Delta t(f)\propto \mathcal{M}^{-5/3}f_\mathrm{GW}^{-8/3}$, provided $\mathcal{M}\sim8\times10^{30}$ kg and $f_\mathrm{GW}\sim1$ Hz. One should note that 1 Hz is below the sensitivity of current-generation GW detectors, which the next-generation detectors (e.g. the Einstein Telescope and LISA) aim to capture. 

Another scenario, using the same example binary system, is the start of observation near merger, where $\beta_C=0.4$, $\rho_C=2G{M}$ and $\widetilde{P}_C\simeq 15$. The pre-merger timelapse until merger is given by the simple reciprocal magnitude of Eq. (\ref{hubble}):
\begin{equation}
\Delta t_\mathrm{pre-merger}=\frac{\rho(t_C)\sqrt{\widetilde{P}(t_C)}}{2\beta(t_C)^5}=\frac{3125\sqrt{15}}{32}GM.
\end{equation}
Thus, for a total mass of $M\sim2\times10^{31}$ kg and given our sample values, the pre-merger timelapse until merger is, naturally, shorter than the previous case, at approximately 18.687 milliseconds. This reflects the near-instantaneous nature of final coalescence, if we begin observation when the binary masses are just about to merge.

\subsubsection{Integral with respect to $\tau'$}

Since Eq. (\ref{hubble}) is observer time-dependent (i.e., if we naively neglect the conversion between observer and conformal times, and treat them as independent), we can treat $H(t)$ as a constant, $H_0$, in the integral over $\tau'$ in Eq. (\ref{2kern}). This is to impose analytical control of an otherwise intractable integral, but the integral ultimately diverges at the asymptote $\tau'=\tau$. To address this, we apply a renormalization approach. Renormalization is commonly used in quantum/effective field theory to handle divergent integrals by isolating and removing unphysical infinities and extracting a physically meaningful, convergent solution.

We regulate the denominator by introducing a small parameter $\varepsilon$, which acts as a cutoff to manage the asymptote near $\tau'\rightarrow\tau$. After obtaining an analytical evaluation, the implied limit of $\varepsilon\rightarrow0$ allows us to Taylor expand the evaluation and remove the divergent terms, keeping only the finite remainder as our physically relevant solution. With the variable substitution $\tau-\tau'=\xi$ and $-d\tau'=d\xi$, the integrand simplifies and evaluates as follows:
\begin{eqnarray}
\int_{0}^{\Delta\tau}\frac{d\xi}{\xi^{4}}~\rightarrow~&&\lim_{\varepsilon\rightarrow0}\int_{0}^{\Delta\tau}\frac{d\xi}{(\xi^2-\varepsilon^2)^{2}}\nonumber\\
&&=\lim_{\varepsilon\rightarrow0}\left[\frac{1}{2\varepsilon^3}\left(\frac{\varepsilon\,\Delta\tau}{\varepsilon^2-\Delta\tau^2}+\mathrm{arctanh}\left(\frac{\Delta\tau}{\varepsilon}\right)\right)\right].
\end{eqnarray}
Here, $\Delta\tau=\tau-\tau_0$. Expanding this result in a Taylor series for small $\varepsilon$ yields 
\begin{eqnarray}
\lim_{\varepsilon\rightarrow0}\Big[\frac{1}{2\varepsilon^3}\Big(\frac{\varepsilon\,\Delta\tau}{\varepsilon^2-\Delta\tau^2}&&+\mathrm{arctanh}\left(\frac{\Delta\tau}{\varepsilon}\right)\Big)\Big]\nonumber\\
&&=-\frac{\pi}{4\varepsilon^2\,\Delta\tau}\sqrt{\frac{-\Delta\tau^2}{\varepsilon^2}}-\frac{1}{3\Delta\tau^3} .
\end{eqnarray}
To obtain the convergent solution, we discard the $\varepsilon^{-3}$ divergent (as well as imaginary) term and keep only the finite, $\varepsilon$-independent term:
\begin{equation}
\int_{0}^{\Delta\tau}\frac{d\xi}{\xi^{4}}~\rightarrow~-\frac{1}{3\Delta\tau^3}.
\end{equation}
Applying this result to Eq. (\ref{2kern}), the original three-fold integral kernel reduces to
\begin{equation} \label{1kern}
\mathcal{K}_3=\frac{4\beta(t)^5}{\rho(t)\sqrt{\widetilde{P}(t)}}\int_0^\infty \frac{d\tau}{a(\tau)}\frac{1}{(\tau-\tau_0)^3}.
\end{equation}

\subsubsection{Integral with respect to $\tau$}

Given that $H_0=\dot{a}/a$, the solution for $a(\tau)$ is an exponential function with a constant Hubble parameter: $a(\tau)=a_0\exp(H_0\tau)$. The initial reference case of $a(\tau_0)=1$ fixes $a_0=1$ and implies $\tau_0=0$, which introduces a divergence in the integral over $\tau$ near $\tau\rightarrow\tau_0=0$. Therefore, another regulation procedure must be done to manage the divergence; this is readily presented by treating $\tau_0$ as our regulating cutoff with the implied limit $\tau_0\rightarrow0$. The regulated integral is evaluated and then Taylor expanded for small $\tau_0$:
\begin{eqnarray}
&&\lim_{\tau_0\rightarrow0}\int_{0}^{\infty}{d\tau}\frac{\exp(-H_0\tau)}{(\tau-\tau_0)^3}\nonumber\\
&&\Rightarrow\left[\frac{1}{2\tau_0^2}+\frac{H_0}{2\tau_0}-\frac{H_0^3\tau_0}{2}-\left(\frac{H_0^2}{2}-\frac{H_0^3\tau_0}{2}\right)\Bigg(\gamma_E+\ln(-H_0\tau_0)\Bigg) \right].
\end{eqnarray}
After discarding the diverging terms, the finite, $\tau_0$-independent remainder is $-H_0^2\gamma_E/2$, with $\gamma_E\simeq0.57722$ being the Euler-Mascheroni constant. Therefore, the three-fold integral kernel is a conformal time-independent factor, however dependent on the observer time:
\begin{equation} \label{0kern}
\mathcal{K}_3=-\frac{8\gamma_E}{\rho(t)^3}\frac{\beta(t)^{15}}{\widetilde{P}(t)^{3/2}}.
\end{equation}
In Eq. (\ref{0kern}), $\rho^3=3V/(4\pi)$ for the CCB mass-shell volume. This establishes that $\mathcal{K}_3\propto V^{-1}$, indicating that quanta dissipation increases and is maximal whenever the volume, respectively, contracts and reaches peak contraction. This, qualitatively, recovers the external classical behavior of CCB-generated GWs with the characteristic enhancement in frequency, amplitude, and wave envelope during coalescence.

\subsection{GW Einstein-Langevin Equation}

 With Eq. (\ref{3kern}) solved analytically as Eq. (\ref{0kern}), we can express the Einstein-Langevin equation for fluctuating gravitons within the contracting volume $V=V(t)$ of a CCB generating GWs:
\begin{equation} \label{eleq3}
\dot{a}=\frac{8\pi\hbar G}{a\,V(t)}\left[\frac{\gamma_E}{9\pi}\,\frac{\beta(t)^{15}}{\widetilde{P}(t)^{3/2}}\,a+\frac{1}{6}\zeta_2(\tau)\right].
\end{equation}
Here, the observer time-dependent dissipation force scales with the loop-quantum-gravity areal quanta $8\pi\hbar G$, suggesting that graviton dissipation may be associated with an underlying quantum nature of spacetime. Given the form of Eq. (\ref{langeq}) when compared to Eq. (\ref{eleq3}), the factor $V/(8\pi\hbar G)$ serves as an effective damping coefficient $\gamma^*$, with the volume $V$ governing the strength of damping in graviton kinematics. As $V$ decreases with contraction, the effective damping effect likewise decreases, leading to increased kinetic intensity of gravitons within the gas.

To introduce conformal time-dependence into the observer time-dependent factors, which would be essential to the Euler iteration scheme when we simulate graviton fluctuations, we can solve $dt=a(\tau)d\tau$ to yield
 \begin{equation} \label{t_to_tau}
 t=\frac{1}{H_0}\exp(H_0\tau)+C_0,
 \end{equation}
where $C_0$ is an integration constant. The mapping of $\tau$, as a result, depends on the log scale when one eventually plots the iterations, inverting the exponential operation applied on $\tau$. We also recall that $H_0$ is negative per Eq. (\ref{hubble}) and treated as a constant with respect to $\tau$. Thus, we state $H_0=-H$ for conceptualization.

We establish initial reference conditions by setting $a(\tau_0)=1$ and $\tau_0=0$. Upon initial reference, the observer's initial time is also null, defining $C_0=1/H$. As $\tau\rightarrow\infty$, $t(\tau\rightarrow\infty)=t_\infty$ converges to $1/H$ due to the exponential decay of $a(\tau)$, indicating a contracting scalar factor. Therefore, complete volume contraction is achieved at coalescence, corresponding to the final observer time of $|H(t)^{-1}|$ discussed in Section \ref{hubbles}.

\subsubsection{The Gaussianity of Graviton Fluctuations}

To interpret Eq. (\ref{eleq3}) as the equation of motion for a single fluctuating graviton, we must introduce the position of the graviton within the CCB mass shell. To do so, we define a graviton's position as $x(\tau)=x_0a(\tau)$, with $x_0$ being the initial position. Applying the discretization procedure offered in Section \ref{discrete}, $\dot{x}(\tau)$ becomes  $\Delta x_i/\Delta\tau$, where $\Delta x_i=x_{i+1}-x_i$. This yields the following Euler scheme for Eq. (\ref{eleq3}):
\begin{equation}\label{eleq4}
x_{i+1}=x_i+\hbar G\left[\left(\frac{2\gamma_E}{3\pi}\right)\frac{\beta(t_i)^{15}}{\widetilde{P}(t_i)^{3/2}\, \rho(t_i)^3}\,a_i+\frac{\theta_{i,2}}{ \rho(t_i)^3\sqrt{\Delta \tau}}\right]\frac{x_0\Delta\tau}{a_i}.
\end{equation}
In Eq. (\ref{eleq4}), each moment $\tau_i$ serves as a timestep, with $\theta_{i,2}$ representing the random noise contribution at each step. This noise simulates the quantum-level ``jitters'' of a graviton moving within a contracting volume. Here, $x_0\simeq10^{-3}$ is chosen for initial simulations, as suggested in Section \ref{discrete}. The factors $\beta$, $\widetilde{P}$, $\rho$ are observer time-dependent, and using Eq. (\ref{t_to_tau}) allows them to be evaluated with respect to conformal time $\tau$. 

Also in Eq. (\ref{eleq4}), it is notable that the dissipation force is generally smaller than the noise term, especially in earlier timesteps. This imbalance suggests that the graviton experiences more ``kicks'' than what equilibration can stabilize, causing an increasingly accumilating deviation out of equilibrium. Consequently, as indicated by the stronger noise contribution, the graviton fluctuations across GW generation is not perfectly Gaussian, arising an issue of inaccuracy should the deviations become largely out of proportion. However, this deviation from Gaussianity can be mitigated by choosing a small $\Delta\tau$ (e.g., $\Delta\tau=10^{-3}$, as suggested in Section \ref{discrete}), ensuring that the system approximates Gaussian behavior. 

Another perspective on Gaussianity involves considering non-equilibirum noise characteristics. The obtained inequality of $F<\sigma$ suggests that graviton noise is akin to $1/f$ noise. This form of noise is known to follow an $\alpha$-stable distribution, where $1<\alpha<2$ is the stability parameter describing deviation from $\alpha=2$ normality \cite{MacKay:2024}. Although setting a small $\Delta\tau$ mitigates large deviations, using an $\alpha$-stable distribution to generate noise kicks offers a realistic alternative for the graviton's non-equilibrium behavior. To retain near-Gaussian behavior, one may set e.g. $\alpha=1.99$ in the simulation. While this may highlight that graviton fluctuations within the CCB mass-shell model exhibit non-Gaussian noise tendencies,  this can be approximated as Gaussian with appropriate timestep adjustments or parameter tuning in the iteration scheme. 

\section{Numerical Results} \label{numres}

The forward Euler scheme for graviton fluctuations is implemented in \textsc{Wolfram Mathematica}, with the length of the conformal-timelapse set to $\mathtt{l=100\,000}$ as suggested in Section \ref{discrete}. Naively at face value, the range in the observer time $t_i\in[0,H_0^{-1}]$ translates to $\tau_i\in[0,\mathtt{l}]$, spanning the entire simulation. At $\tau_l=\mathtt{l}$, the coupling $H_0t_\infty=1$ defines the moment of merger, which corresponds to $H_0\tau_l=\ln(2)$ via Eq. (\ref{t_to_tau}). To account for the exponential decay of the scalar factor due to a negative $H_0$, the observer time and scalar factor are defined as follows, i.e. independent of the explicit definition of $H_0$ via Eq. (\ref{hubble}):
\begin{verbatim}
t[i_]:=-l(Exp[-Log[2]*i/l]-1)/Log[2],
a[i_]:=Exp[-Log[2]*i/l]
 \end{verbatim}
 
The observer time-dependent parameters $\beta$, $\widetilde{P}$, and $\rho$ evolve respectively as $\beta(t)\rightarrow0.435$ (the speed limit imposed by the CCB mass-shell model), $\widetilde{P}(t)\rightarrow15$, and $\rho(t)\rightarrow2G{M}$ towards merger, as $\tau_i\rightarrow\mathtt{l}$. With $\mathtt{t[i]}$ as the observer time, these evolutions are modeled heuristically as
 \begin{verbatim}
beta[t_]:=0.435 * E*t/(2*l)
P[t_]:=15-9Exp[-25(t/l)^5]
rho[t_]:=2(1+Sqrt[1-1.95(t/l)^2])
 \end{verbatim}
  
It is worth repeating that the parameters above are not uniquely defined by any physical definition, but rather expressed as phenomenological test functions to capture the expected dynamics of the system. E.g., the semi-latus rectum $\mathtt{P[t[i]]}$ models eccentricity dependence, transitioning from $\widetilde{P}\simeq6$ for low eccentricities to $\widetilde{P}\simeq15$ for high eccentricities as time elapses. This eccentricity-dependent phenomenology is given by the $\mathtt{(t[i]/l)^5}$ factorization in an exponential decay, demonstrating a strong cut-off. Similarly, the dimensionless radius of the CCB mass shell, $\mathtt{rho[t[i]]}$, is modeled after a contracting Kerr outer radius, emphasizing volumetric contraction throughout coalescence.

To simulate noise, random kicks are generated using $\mathtt{RandomVariate}$ and $\mathtt{NormalDistribution[0,1]}$. Alternatively, near-Gaussian $\alpha$-stable noise is modeled with $\mathtt{StableDistribution[alpha, 0, 0, 1]}$ with e.g. $\mathtt{alpha=1.99}$. This flexibility allows exploration of non-Gaussian noise dynamics, however this work focuses on the normally-distributed Gaussian iterations. To manage the extreme scaling of the effective damping coefficient (recovering $c$) $\gamma^{*}=G^2{M}^3/(\hbar c^3)\sim10^{81}\,\mathrm{m}$ for a total mass of $M\sim10^{31}\,\mathrm{kg}$ -- making the reciprocal extremely small, filtering out any visible iterations --, the iteration is rescaled in units of $1/\gamma^*\sim10^{-81}\,\mathrm{m^{-1}}$, in order to ensure numerical stability. To achieve a classical, GR-determinable strain amplitude of order $A_0\sim10^{-21}$ via Eqs. (\ref{gwsourced}) and (\ref{gwvacuum}), a coherent-state abundance number of order $N_c\sim10^{63}$ is needed, such that $A_0\approx N_c\, x_0/\gamma^*$ with $x_0\sim10^{-3}\,\mathrm{m}$ via Eq. (\ref{eleq4}). 

For Gaussian noise generation, the simulation parameters are defined and presented as follows:
 \begin{verbatim}
dtau = 10^(-3);

force = Table[0.1225 *beta[t[i]]^(15)/(rho[t[i]]P[t[i]]^(1/2))^3, 
                          {i, l}];
data = RandomVariate[NormalDistribution[0, 1], l];
kicks = Table[(data[[i]]/((rho[t[i]])^3 *dtau^(1/2))), {i, l}];

x0 = 10^(-3);
x[1] = x0;
 \end{verbatim}
The last two lines define the initial position of the graviton. The iteration loop is structured by a central \texttt{Do} command with embedded \texttt{If} conditions:
 \begin{verbatim} 
Do[dx=(force[[i]]*a[i]+kicks[[i]])x0*dtau/a[i];
    If[x[i]==x0 && kicks[[i]]<= force[[i]],dx=0];
    x[i+1]=x[i]+dx;
    If[x[i+1]<0,x[i+1]=0], {i,1,l}];

iterations=Table[x[j], {j,l}];
 \end{verbatim}
The first $\mathtt{If}$ condition encourages fluctuations from $x_0$ while preventing deviations that collapse the simulation into a flat-line for all timesteps. The second $\mathtt{If}$ condition enforces $x_{i+1}>0$, consistent with the barrier-linear potential well.
 
\begin{figure}[h!]
\centering
\includegraphics[width=0.9\textwidth]{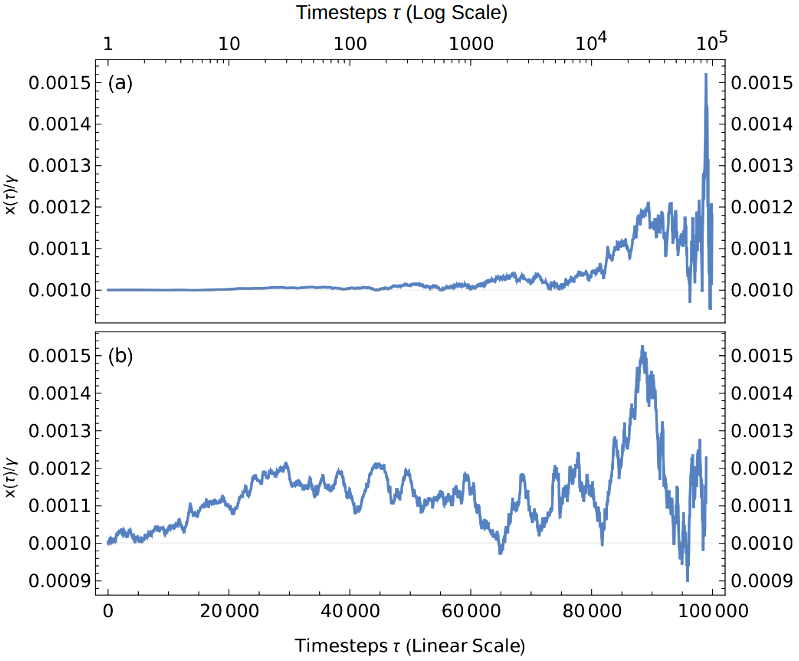}
\caption{\label{fig:gravits} A graviton fluctuation simulation under the Gaussian distribution. Fluctuations in the ``strain" (i.e. the unitless $x(\tau)/\gamma^*$ in blue) are shown under two plotting scales: panel (a) plots the iterations in the log-linear scale, and panel (b) plots the same iterations in the default linear-linear scale.}
\end{figure}

Over the range of $\mathtt{i\in[1,l]}$, both the force and kick size increase due to radial contraction, aligning with the expected astrophysical dynamics of GW generation. By interpolating the $\mathtt{iterations}$ data points into a continuous function $x(\tau)$ and plotting it against the timesteps $\tau$ using $\mathtt{LogLinearPlot}$ -- such that the log scale of the $\tau$-axis inverts the inherent exponential scaling --, we yield a fluctuation pattern akin to those shown in Figure \ref{fig:gravits}(a) under Gaussian noise. For comparison, Figure \ref{fig:gravits}(b) depicts the same, respective noise realizations using the default $\mathtt{Plot}$ command (the linear-scale in the $\tau$-axis), where only iterations at timesteps on the order of $\sim10^4$ are visually distinguishable. 
 
\section{Discussion} \label{disc}

The simulations depicted in Figure \ref{fig:gravits} illustrate how these heuristic inputs lead to emergent behaviors that can, with further refinements, provide deeper insights into the role of graviton fluctuations in GW generation. The choice of a log-linear representation is of particular relevance, as it not only enables the visualization of early-time fluctuations, but also captures the characteristic chirp rise in the inspiral-merger phases.  While the results contain features that resemble a GW signal as a Wiener process, these findings should be interpreted as a proof-of-concept rather than a definitive prediction. Further refinement is possible by adjusting key parameters and the compuational algorithm, redefining the heuristic test functions for the observer-time parameters in the iteration code, testing alternative noise distributions, or revisiting the analytical derivation of the dissipation kernel.

Figure \ref{fig:gravits}(a) illustrates fluctuations bearing a qualitative (if not exact) resemblance to macroscopic gravitational waveforms, particularly the build-up of sinusoidal-like perturbations at later timesteps, and a sharp peak near the final pulse. This final ``kick'' in the simulation corresponds to an abrupt increase in graviton fluctuations, reflecting an intensified stochastic effect as coalescence approaches. In the context of GW astrophysics, this behavior mirrors the chirp rise in emitted GW waveforms, which are seen by Earth-based (i.e. very distant) observers. Furthermore, one may claim rather daringly that the log-scale in $\tau$ for an up-close observer -- near the contracting CCB mass-shell -- is equivalent to the linear-scale in $t$ for a very-distant observer. This reconciliation intends to equate these relative observations of the same event, even if individually measured in a different mapping or scaling, which ensures that the same indisputable physics is captured. 

It should be noted that the inherent randomness of noise generation precludes exact figure reporduction, Nonetheless, the essential waveform characteristics remain consistent, underscoring how increased graviton kinematics in the ``microscopic" level correlate to macroscopic coalescence dynamics. 

\subsection{Analytical Waveform Matching}

It is of interest to investigate whether the iterated Wiener process profile seen in Figure \ref{fig:gravits}, which visually compares to an inspiral-merger waveform, is an effective mirror of the CCB mass-shell waveforms described as e.g. Eqs. (\ref{hp2}) and (\ref{hc2}) (which only map the inspiral-merger phases). This is done with pen and paper by considering how $\Omega,~\dot\Omega$, and $\ddot{\Omega}$ are defined (here $^\bullet=d/dt$ temporarily), whether we begin with a heuristic Ansatz for $\Omega=\Omega(t)$ and differentiate up to second order, or begin with an Ansatz for $\ddot{\Omega}$ and yield lower order expressions from integration. 

To provide a representative example, one may choose the following Ansatz $\ddot{\Omega}=\epsilon$, analogously defining a constant acceleration in the orbital frequency. Similar to finding the system of kinematic equations in fundamental mechanics, we define $\dot{\Omega}$ and $\Omega$ via integration from the initial ``acceleration" gauge:
\numparts
\begin{eqnarray}
& \dot{\Omega}=\epsilon t+\omega,\\\label{om}
&\Omega=\frac{1}{2}\epsilon t^2+\omega t+\Omega_0.
\end{eqnarray}
\endnumparts
One can see via Eq. (\ref{om}) that the quadratic term $\epsilon t^2/2$ serves as the leading contributor to the quadratic-in-behavior chirp rise in the normalized frequency enhancement $f_\mathrm{GW}\propto2\Omega$, as $t\rightarrow t_C$.

The provided forms of $\dot{\Omega}$ and $\Omega$ look Newtonian in this sense, which couples with the intuition that the acceleration of the CCB is due to a characteristically-Newtonian inward attraction. One should keep in mind that the orbital phase $\varphi\propto\Omega t$ of the GW is conventionally PN-expanded, where the expansion follows by the leading orders in $v^2/c^2\sim GM/L$ with $L$ being the binary seperation. In Eq. (\ref{om}), an expansion arises in leading orders of the dynamic observer time $t$, i.e. $2\Omega t= a_3t^3+a_2t^2+a_1t$, where respectively $a_3=\epsilon$, $a_2=2\omega$, and $a_1=2\Omega_0$.

Given these Newtonian-like expressions for the rates of change in $\Omega$, and for $\Omega$ itself, the polarization envelope functions $\mathfrak{E}_{+/\times}$ that are specific to Eqs. (\ref{hp2}) and (\ref{hc2}) are defined as follows (under the assumption $\Omega_0\sim0$ for a negligibly small, initally weak orbital frequency):
\numparts
\begin{eqnarray}
&&\mathfrak{E}^\mathrm{sin}_+=\mathfrak{E}^\mathrm{cos}_\times= -\frac{(\epsilon t+\omega)}{(\epsilon t+2\omega)}(3\epsilon t+4\omega)+\omega+\frac{3}{2}\epsilon t,\\
&&\mathfrak{E}^\mathrm{sin}_\times=-\mathfrak{E}^\mathrm{cos}_+=\frac{1}{4}\left(-\frac{2\epsilon}{\epsilon t^2+2\omega t}+\frac{4(\epsilon t+\omega)^2}{(\epsilon t^2+\omega t)^2} \right)-\left(\frac{3}{2}\epsilon t^2+2\omega t \right)^2.
\end{eqnarray}
\endnumparts
In addition, the sinusoidal profiles $h_{+/\times}\propto\exp(i\,2\Omega t)$ are revised such that the argument is the polynomial in $t$: $2\Omega t = \epsilon t^3+2\omega t^2$.

With two tunable gauge parameters $\epsilon$ and $\omega$ (as well as the third gauge parameter $\Omega_0$, if one chooses not to neglect it), one would calibrate these values to match a specific event's frequency-time plot inferred in the open-access \textit{GWOSC} catalog \cite{GWOSC}. This is while an analytical expression for each gauge parameter in terms of CCB observables is lacking. However, in this example, we explore the asymptotic behavior of ``strong $\omega$" and ``strong $\epsilon$". This is to qualitatively explain the behavior of the waveforms under such asymptotic regimes, and how they interplay concurrently in the case of $\epsilon \sim \omega$.

For ``strong $\omega$" (alternatively a weak or non-existing $\epsilon$), such that the orbital frequency increases linearly -- applicible to nearly-circular inspiral phases --,
\numparts
\begin{eqnarray}
&\mathfrak{E}^\mathrm{sin}_+=\mathfrak{E}^\mathrm{cos}_\times\simeq -\omega,\\
&\mathfrak{E}^\mathrm{sin}_\times=-\mathfrak{E}^\mathrm{cos}_+\simeq\frac{1}{t^2} -4\omega^2 t^2\approx  -4\omega^2 t^2,
\end{eqnarray}
\endnumparts
and the waveforms $h_{+/\times}\propto\exp(i\,2\Omega t)$ are internally influenced by the overpowering $\omega$ term: $2\Omega t \simeq 2\omega t^2$. On the other hand, for ``strong $\epsilon$" (alternatively a relatively-weak $\omega$) such that the orbital frequency increases quadratically -- more relevant in the inspiral-merger transition --,
\numparts
\begin{eqnarray}
&\mathfrak{E}^\mathrm{sin}_+=\mathfrak{E}^\mathrm{cos}_\times\simeq -\frac{3}{2}\epsilon t,\\
&\mathfrak{E}^\mathrm{sin}_\times=-\mathfrak{E}^\mathrm{cos}_+\simeq\frac{1}{2t^2} -\frac{9}{4}\epsilon^2 t^4 \approx -\frac{9}{4}\epsilon^2 t^4,
\end{eqnarray}
\endnumparts
and the waveforms $h_{+/\times}\propto\exp(i\,2\Omega t)$ are internally influenced by the overpowering $\epsilon$ term: $2\Omega t \simeq \epsilon t^3$. 

In the internal behavior of the sinusoidal profiles, both regimes illustrate a progressively-tightening waveform structure, such that each cycle becomes progressively narrower with an ever-decreasing wavelength. In comparison, the ``strong $\epsilon$" case demonstates that this wave-tightening occurs much sooner than the ``strong $\omega$" case. In addition, the wave envelope evolution in both regimes are both time-forward enhancing but behaviorally different. In the ``strong $\omega$" case, this envelope enhancement is drawn almost linearly, while the ``strong $\epsilon$" case draws a ``Gabriel's Horn" shape. If these cases are mutually interplayed, these mechanisms are consistent with the conventional viewpoint that the wave cycles and amplitude/envelope enhancement in GW profiles, respectively, ``tighten" and are quadratically scaled due to the chirp rise towards merger. These qualitative descriptions are furthermore demonstrated in the iterative approach depicted in Figure \ref{fig:gravits}.

\subsection{Further Justification for the Brownian Graviton Bath} \label{shmear}

\subsubsection{On Gravitational EFTs}

Using the Einstein-Langevin equation to model graviton fluctuations in the context of GW generation is consistent with the theoretical interpretation of classical GWs as coherent-state gravitons. While they have not been detected, this correspondence has served as a useful analytical tool when approaching relativistic gravity as an effective field theory (EFT). As a representative example, the worldline quantum field theory (WQFT) framework \cite{Mogull:2020sak} formulates the early inspiral phase as a perturbative $2\rightarrow2$ scattering problem between compact bodies. These scattering events -- provided the background metric is asymptotically Minkowskian -- can be drawn as an assortment of Feynman diagrams, using first-quantization principles to quantize the path deflection of the worldlines $z_i^\mu$ and the first-order metric perturbations $h_{\mu\nu}$. From these quantizations, Feynman rules were derived in order to construct diagrams and calculate their corresponding amplitudes. Like other gravitational EFTs \cite{Feynman:2002, Goldberger:2004jt, Goldberger:2006bd, Goldberger:2007hy, Kol:2007bc, Goldberger:2009qd, Foffa:2013qca, Rothstein:2014sra, Porto:2016pyg, Levi:2018nxp, Rafie-Zinedine:2018izq, Mogull:2020sak, Aoki:2024boe, Blas:2020och, Delgado:2022uzu, Herrero-Valea:2022lfd}, the coupling constant at one vertex is gauged by $\kappa\sim\sqrt{G}$. Thus, the PM order $m$ is gauged by $G^m\sim\kappa^{2m}$, following conventional PM expansion. 

In a standard QFT, e.g. QED and QCD, the total amplitude of a scattering event typically follows a polynomial expansion in the coupling constant-squared (e.g. $g_e^2\propto\alpha_e\simeq1/137$ for QED and $g_s^2\propto\alpha_s\simeq1$ for QCD). These polynomial orders stem from perturbative corrections to the very basic, ``tree-level" Feynman diagram, with each diagram variation going by the starting or a higher perturbation order. In gravitational EFTs such as WQFT, the inspiral phase is described at tree-level as a 1PM-order graviton exchange between two scalar worldlines. Higher PM orders appear pictorally as increasingly complex diagrams, which often involve e.g. internal graviton loops and radiation branches. 

Up to the 5PM order in WQFT (where 5PM diagrams contain 10 vertices in one diagram), the total amplitude calculation requires evaluating at least 417 variational diagrams, often relying on computational software to decompose master integrals and e.g. using a summation approximation to evaluate such integrations effectively. While these calculations are state-of-the-art and they agree well with numerical relativity for $L/GM>14$ \cite{Driesse:2024feo}, the rapid growth in diagrammatic complexity puts computational expense into consideration, and thus motivates the exploration of a complementary alternative (even if an oversimplifying approximation). 

Should one extend up to e.g. the 6PM order, where individual 6PM variations involve 12 vertices and the number of total diagrams grows rapidly, it becomes natural to consider an alternative where one ``smears" the internal graviton loops and branches into a statistical, ideal gas analogy (i.e. a Brownian graviton bath). In this view, the complicated graviton-graviton interactions occuring between two scalar worldlines (e.g. within a CCB hollow mass-shell) are treated in an effective kinetic-theoretical sense rather than explicitly via diagrams. Therefore, our focus is shifted from scattering worldlines to scattering gravitons, as long as they are confined in the CCB mass-shell.   

\subsubsection{Graviton-Graviton Total Cross Section}

 Within this statistical framework, where the graviton sector is treated as an effective ideal gas, the total amplitude associated with tree-level graviton-graviton scattering, together with its leading variations, becomes more relevant than individual high-PM, diagram-resolved contributions. The focus on the tree-level, 1PM diagrams is analogous to asymptotic freedom in QCD \cite{Gross:1973id}, in which the ultra-relativistic kinematics of quarks and gluons screens the strong coupling to be small at high-energy scales. As a QFT with a large coupling gauge, the higher perturbation orders become more probable than the basic interactions (unlike in QED); this parallels with higher PM orders and their effectiveness in e.g. WQFT and standard PM calculations. However, asymptotic freedom constrains the strong coupling to be small, allowing the tree-level diagrams to take kinematic precedence. Therefore, we adopt this for our high-energy gravitons in the Brownian bath, as the coalescing binary dynamics serve as the kinetic gauge-screening to graviton self-interactions and as a source of inherent pressurization from a contracting volume (see footnote\footnote{The prospect of a kinematically-screened, gravitational gauge coupling $\kappa\rightarrow\kappa(q)$ is analogous to the stochastic corrections in the EFEs in stochastic gravity theory, see Section \ref{sect:braun}. These extensions essentially rest in (albeit ad hoc) randomness-inducing expansions in the gravitational constant $G\rightarrow G'=G+\sigma\zeta_2$.}). 

This allows for the resulting total amplitude from such graviton-graviton interactions to compute the differential cross section $d\sigma/d\Omega$ (equivalently $d\sigma/d\hat{t}$ in particle physics literature, where $\hat{t}$ denotes momentum transfer). From the total cross section $\sigma=\int (d\sigma/d\Omega)d\Omega=\int (d\sigma/d\hat{t})d\hat{t}$, one may yield an effective impact parameter $b$ via $\sigma=\pi b^2$. In WQFT, this impact parameter corresponds to the CCB seperation $L$ given worldline scatterings, but for an isolated graviton-graviton interaction, this impact parameter would be extremely small in comparison. However, we leverage the statistical treatment of the gravitons in a Brownian bath, integrating over a distribution of scattering events and accessible center-of-momentum energies. The resulting impact parameter would therefore be enhanced relative to that single encounter, such that $b/(GM)$ may be relevant in this context. 

In a graviton-rich ideal gas, graviton-graviton scatterings are assumed to be ubiquitous. To fix notation and interaction strength, we adopt the Feynman rules of Ref. \cite{Rafie-Zinedine:2018izq}, for which the gauge coupling is defined as $\kappa=\sqrt{8\pi G/\hbar}$. Equivalently, this may be expressed in terms of the Planck mass: $\kappa=\sqrt{8\pi}/m_P$ \cite{Delgado:2022uzu, Herrero-Valea:2022lfd}.  At tree level \cite{Rafie-Zinedine:2018izq, Delgado:2022uzu, Herrero-Valea:2022lfd}, the graviton-graviton scattering amplitude accounts for relevant exchange channels and the $\pm2$ helicity permutations of the external particle lines:
\begin{equation} \label{totamp}
\mathcal{A}_{\mathrm{tot}}=\kappa^2\left(\frac{\hat{s}^3}{\hat{t}\hat{u}}+\frac{\hat{u}^3}{\hat{s}\hat{t}}+\frac{\hat{t}^3}{\hat{s}\hat{u}} \right).
\end{equation}
In the above, $\hat{s},~\hat{t},~\hat{u}$ are the standard Mandelstam variables used in Feynamn-diagram calculations ($\sqrt{\hat{s}}$ corresponds to center-of-momentum energy; while both $\sqrt{\hat{t}}$ and $\sqrt{\hat{u}}$ are associated with momentum transfer). For massless particles such as gravitons, these variables satisfy $\hat{u}=-\hat{t}-\hat{s}$, such that Eq. (\ref{totamp}) depends only on $\hat{s}$ and $\hat{t}$. As is well known, graviton-graviton scattering exhibits an infrared divergence in the low-energy regime. In the present case, this arises from the $\hat{t}^3/(\hat{s}\hat{u})$ channel for small $\hat{s}$, with $\hat{u}=-\hat{t}-\hat{s}$ readily implied \cite{Herrero-Valea:2022lfd}. 

In the context of the graviton sector smeared into an ideal gas, we are interested in the graviton modes that contribute coherently to macroscopic metric perturbations. In such a setting, the relevant interactions are effectively dominated by high center-of-momentum energies, which correspond instead to large $\hat{s}$. This motivates considering the asymptotic high-energy limit of the tree-level amplitude as an effective description, without attempting to resolve infrared-sensitive dynamics. Accordingly, Eq. (\ref{totamp}) may be approximated in the limit $\hat{s}\rightarrow\infty$, with an additional 1/2 included to correct for the over-counting of identical exchange channels for same-species particles \cite{Delgado:2022uzu}:
\begin{equation} \label{totamphe}
\lim_{\hat{s}\rightarrow\infty}\mathcal{A}_{\mathrm{tot}}\equiv\mathcal{A}_{\mathrm{he}}=-\frac{8\pi G}{\hbar}  \frac{\hat{s}^2}{\hat{t}}.
\end{equation} 

 Differential cross sections are proportional to the square of the amplitude, readily summed over final-state spin helicities via Casimir's trick and averaged over the initial-state spin values: $d\sigma/d\hat{t}\propto\langle|\mathcal{A}_{\mathrm{tot}}|^2\rangle$ \cite{griffiths}. For the high-energy amplitude in Eq. (\ref{totamphe}), the squared magnitude is averaged over the spin values of each incoming graviton, yielding a normalization factor of $2^2=4$. One therefore finds, rather simply: 
\begin{equation}
\langle|\mathcal{A}_{\mathrm{he}}|^2\rangle=\frac{16\pi^2 G^2}{\hbar^2}  \frac{\hat{s}^4}{\hat{t}^2}.
\end{equation} 
This leads to the differential cross section:
\begin{equation}
\frac{d\sigma}{d\hat{t}}\equiv\frac{\hbar^2}{16\pi \hat{s}^2}\langle|\mathcal{A}_{\mathrm{he}}|^2\rangle={\pi} G^2 \frac{\hat{s}^2}{\hat{t}^2},
\end{equation} 
as well as an estimate of the total cross section by introducing a lower cutoff at $\hat{t}\sim-\hat{s}$ appropriate to the effective high-energy regime,
\begin{equation}
\sigma_{\mathrm{he}}\simeq\int_{-\hat{s}}\frac{d\sigma}{d\hat{t}}d\hat{t}={\pi} G^2 \hat{s}.
\end{equation} 

Since the total cross section depends on the center-of-momentum energy variable $\hat{s}$, it is useful within a statistical description to consider an average value. This removes the explicit energy dependence in favor of a characteristic scale, which in typical statistical contexts is thermal. For a graviton ideal gas whose dynamics are not thermal, we consider an effective thermal energy which we will shortly determine. As the average depends on the distribution functions of the interacting particles, and gravitons are massless bosons, the interacting gravitons obey the Bose-Einstein (BE) distribution; the effective thermal average of the high-energy cross section becomes 
\begin{equation}
\langle \sigma_{\mathrm{he}}\rangle= {\pi} G^2\langle \hat{s}\rangle=14.5927\pi\, G^2\Theta^2,
\end{equation}
where $\Theta$ denotes an effective thermal energy scale. The derivation of $\langle \hat{s}\rangle$ under BE statistics is provided in \ref{avgs}. 

As discussed in Section \ref{sect:braun}, the stochasticity of graviton fluctuations within the CCB mass-shell is induced in part by excitations sourced by the inspiraling binary bodies. Thus, this graviton stochastiticy may be parameterized by a characteristic energy scale that is reflective of its macroscopic source. In other words, this effective energy scale is set by the surface energy of the CCB mass-shell across coalescence, which gauges the randomness of metric fluctuations within, and naturally outside, the hollow shell. Under this identificiation, the average total cross section inherits its dependence on the surface energy through Eq. (\ref{endens1}), where the total energy is given by $E=T_{00}V$ evaluated at $r=\rho$. Approximating the numerical factor $14.5927\approx15$, one finds:
\begin{equation}
\langle \sigma_{\mathrm{he}}\rangle\approx \frac{5\pi^3}{3}\,\frac{G^4\mu^4}{\rho^2}\left(1.577-8.320 \frac{a^2}{\rho^2}\right)^2.
\end{equation}
The high-energy regime relevant to the Brownian graviton picture corresponds to late inspiral and near-merger phases, where graviton excitations are strongly sourced and the effective stochastic description becomes most appropriate. It is worth noting that this regime lies beyond the domain of the WQFT framework, which assumes an asymptotically Minkowskian background and therefore does not directly describe near-merger kinematics. In this high-energy sense, the graviton-graviton impact parameter serves as an effective interaction scale that, already given statistical averaging, yields a macroscopic interpretation relevant to peak GW emission. 

More specifically, it is of interest to determine if the impact parameter is smaller than the CCB mass-shell radius, such that the self-interacting gravitons are within the confines of the hollow shell across all physically-relevant phases. Thus, we consider a mass-shell radius $\rho$ that is of order of $GM$: $\rho=qGM$, where $q$ is a scaling factor that is e.g. equal to 2 at time $t=t_C$ (and therefore larger at earlier times). We yield, approximating $1.577\approx1.6$ and defining $a/\rho=\beta$:
\begin{eqnarray}
\Rightarrow\quad \langle \sigma_{\mathrm{he}}\rangle&\approx \frac{8\pi^3}{3}\,\frac{G^2\mu^4}{q^2M^2}\left(1-\frac{26}{5}\beta^2 \right)^2\nonumber\\
&=\frac{8\pi^3}{3}\,\frac{\nu^2}{q^2}G^2\mu^2\left(1-\frac{26}{5}\beta^2 \right)^2,
\end{eqnarray}
where $\nu=\mu/M$ is the symmetric mass ratio. We can now equate the averaged high-energy total cross section to $\langle \sigma\rangle=\pi \langle b^2\rangle$, where we find the root-mean-square value for the impact parameter $\sqrt{\langle b^2\rangle}=b_\mathrm{rms}$ and its scale $b_\mathrm{rms}/(GM)$:
\begin{eqnarray}
\Rightarrow\quad&    \sqrt{\langle b^2\rangle}\equiv b_\mathrm{rms} =\pi\sqrt{\frac{8}{3}}\,\frac{\nu}{q}G\mu\left(1-\frac{26}{5}\beta^2 \right),\\\label{bgm}
& \frac{ b_\mathrm{rms}}{GM}=\pi\sqrt{\frac{8}{3}}\,\frac{\nu^2}{q}\left(1-\frac{26}{5}\beta^2 \right).
\end{eqnarray}
As we are assuming equal mass binaries, as we had in Section \ref{hubbles}, $\nu=1/4$. 

We can infer via Eq. (\ref{bgm}) that a large value of $q$ and a small value of $\beta$ -- which are relevant to (nearly-)circular orbits and early inspiral -- produces a very miniscule and negligible root-mean-square impact parameter. Therefore, early-stage graviton self-interactions (i.e. early-stage GW emissions) are too miniscule to yield any meaningful measurement. For e.g. early inspiral where $\beta=0.1$ and we suppose $q=15$, we calculate a root-mean-square impact parameter of $b_\mathrm{rms}/(GM)\simeq 0.0203$. And for the near-merger case where $\beta=0.4$ (close to the model's speed-ratio limit) and $q=2$, we calculate $b_\mathrm{rms}/(GM)\simeq0.0269$. 

These values are well within the CCB mass-shell, and we build a range of values of $b_\mathrm{rms}/(GM)\in(0,\,0.0269]$. In other words, high-energy graviton-graviton collisions in the Brownian sense take place in the CCB mass-shell interior, and the mass-shell's morphology affects the effective measurability of their cross section. The measurability of graviton interactions, and the resulting randomness in the jitters they track e.g. as a Wiener process, is also demonstrated in the iterative approach depicted in Figure \ref{fig:gravits}.

\section{Concluding Statements}\label{concl}

In this study, we applied the Einstein-Langevin equation to the interior of a CCB mass-shell model with a contracting volume. Our primary motivation was to explore whether the enclosed first-order metric perturbations, modeled as graviton fluctuations, could exhibit stochastic behavior analogous to Brownian motion during coalescence. To enable numerical analysis, we reformulated the original second-order differential equation (Eq. [\ref{eleq}]) into a first-order Langevin-like equation, simplifying the role of the dissipation kernel as an effectively constant dissipation force. 

To establish a connection between this stochastic approach and GW generation, Section \ref{quantder} introduced a Hubble-like parameter specific to compact binary coalescence, incorporating dependencies on observer time-dependent osculating eccentricity, semi-latus rectum, and the radius of the CCB mass shell. Through renormalization-based regularization schemes, we extracted a non-zero solution from otherwise divergent integrals. This enables a discretized numerical simulation of graviton fluctuations, as detailed in Section \ref{numres}.

Future extensions to this work both lie in the computational and analytical parts. Computationally, it is of interest to perfect the introduced numerical scheme, e.g. by implementing better phenomenological test functions for the observer-time measurables, as well as further computational implementation to ensure a Wiener process output that is comparable to macroscopic GW signals. In the analytical effort, it is of interest to continue the discussion topics as their own studies. These topics of discussion, as listed in Section \ref{disc}, involve a graviton ideal gas inside a CCB mass-shell and engaging in EFT approaches to late-stage and post-merger gravitational radiation. For instance, there is an interest to e.g. calculate the cross section of LIGO-like scalar-graviton Compton scatterings -- which intends to assist in the pursuit to detect gravitons -- (see Ref. \cite{MacKay:2024sgw}), as well as heuristic, background-dependent EFT techniques to complement e.g. WQFT for late-stage, pre-merger GW emissions.



\section*{Statement Declarations}

\subsection*{Conflict of Interest}
The author declares no conflicts of interest.

\subsection*{Data Access Statement}
As a theoretical study, this work generates no original data. Data from cited LIGO observations are publicly available.

\subsection*{Ethics Statement}
No ethical issues arise, as no test subjects are involved. This paper adheres to academic integrity.

\subsection*{Funding Statement}
This work received no funding.


\appendix

\section{Deriving $\langle \hat{s}\rangle$ under Bose-Einstein Statistics} \label{avgs}

In Feynman diagram calculations, the $\hat{s}$-Mandelstam variable is defined as the square of the sum between the two incoming (or alternatively outgoing) 4-momenta of the external particle lines: $\hat{s}=(p_1+p_2)^2=(p_3+p_4)^2$. Conveniently, it is Lorentz-invariant; in the gas rest frame, the $\hat{s}$-Mandelstam variable is expanded and specifically defined for massless particles:
\begin{equation}
\hat{s}=p_1^2+p_2^2+2p_1\cdot p_2=2(E_1E_2-\vec{p}_1\cdot\vec{p}_2).
\end{equation} 
In the above, $p_i^2=m_i^2$ via the 4-momentum product, which is equal to zero for massless particles. This leads to $E_i=|\vec{p}_i|$. In the gas rest frame, the 3-momenta are projected by their respective azimuthal and polar angles, where the dot product is taken in spherical coordinates:
\begin{equation}
\vec{p}_1\cdot\vec{p}_2=|\vec{p}_1||\vec{p}_2|\left(\cos(\phi_1-\phi_2)\sin\theta_1\sin\theta_2+\cos\theta_1\cos\theta_2 \right).
\end{equation}  
This defines the $\hat{s}$-Mandelstam variable for massless particles in the gas rest frame as
\begin{equation} \label{sexp}
\hat{s}=2|\vec{p}_1||\vec{p}_2|(1-\cos(\phi_1-\phi_2)\sin\theta_1\sin\theta_2-\cos\theta_1\cos\theta_2).
\end{equation} 

The thermal average of any quantity, such as $\hat{s}$, is defined as a two-distribution average:
\begin{eqnarray}
&\langle \hat{s}\rangle=\frac{1}{Z_1Z_2}\int_{-\infty}^{\infty} d^3\vec{p}_1d^3\vec{p}_2\,\hat{s}\,f_1(\vec{p}_1)f_2(\vec{p}_2),\\
&Z_i=\int_{-\infty}^\infty d^3\vec{p}_i\,f_i(\vec{p}_i),
\end{eqnarray}
where $f(\vec{p}_i)=\{\exp(E(\vec{p}_i)/\Theta)\pm1\}^{-1}$ under quantum statistics, with $\Theta$ conventially being a thermal energy scale $k_BT$. In general, this applies for any combination of two distribution functions, depending on the particles that are involved in the given interaction. For two massless bosons, such as gravitons, both distribution functions are the massless Bose-Einstein (BE) distribution with $f(\vec{p}_i)=\{\exp(|\vec{p}_i|/\Theta)-1\}^{-1}$. Thus, $Z_1Z_2=(Z_0)^2$, where $Z_0=8\pi\zeta(3)\Theta^3$ is the normalization constant for the massless BE distribution.

The distribution functions, themselves, are only momentum-dependent, with no angular dependency. Via Eq. (\ref{sexp}) and defining $d^3\vec{p}_i=|\vec{p}_i|^2d|\vec{p}_i|\sin\theta_id\theta_id\phi_i$, the 2-integral can be separated between the momentum part and the angular part:
\begin{eqnarray}
&\langle \hat{s}\rangle=\frac{2}{(8\pi\zeta(3)\Theta^3)^2}I_{\mathrm{mom.}}I_{\mathrm{ang.}},\\
&\mathrm{where}~~~I_{\mathrm{mom.}}=\int_{0}^{\infty} d|\vec{p}_1|d|\vec{p}_2|\,|\vec{p}_1|^3|\vec{p}_2|^3\,f_1(\vec{p}_1\!)f_2(\vec{p}_2\!)=\frac{\pi^8}{225}\Theta^8\\
&\mathrm{and}~~~I_{\mathrm{ang.}}=\int_{0}^\pi d\theta_1d\theta_2\int_{0}^{2\pi}d\phi_1d\phi_2\,\sin\theta_1\sin\theta_2\nonumber\\
&\quad\quad\quad\times (1-\cos(\phi_1-\phi_2)\sin\theta_1\sin\theta_2-\cos\theta_1\cos\theta_2)=16\pi^2.
\end{eqnarray}
Therefore, the thermal average of the $\hat{s}$-Mandelstam variable under massless BE statistics is
\begin{equation}
\langle \hat{s}\rangle=\frac{\pi^8}{450\zeta(3)^2}\Theta^2=14.5927\,\Theta^2.
\end{equation}





\section*{References}

\begin{thebibliography}{99}
\bibitem{GWOSC}
Gravitational Wave Open Science Center ({https://gwosc.org/eventapi/html/GWTC/})

\bibitem{LIGOScientific:2018mvr}
Abbott~B~P, LIGO Scientific and Virgo,~ \textit{et al.} 2019 \textit{
Phys. Rev. X} \textbf{9}, no.3, 031040

\bibitem{LIGOScientific:2021usb}
Abbott~R, LIGO Scientific and VIRGO,~\textit{et al.} 2024 \textit{
Phys. Rev. D} \textbf{109}, no.2, 022001

\bibitem{KAGRA:2021vkt}
Abbott~R, KAGRA, VIRGO and LIGO Scientific, \textit{et al.} 2023 \textit{
Phys. Rev. X} \textbf{13}, no.4, 041039

\bibitem{LIGOScientific:2025slb}
Abac~A~G, LIGO Scientific, VIRGO and KAGRA,~\textit{et al.} 
[arXiv:2508.18082 [gr-qc]].


\bibitem{Blanchet:2013haa}
Blanchet~L 2014 \textit{
Living Rev. Rel.} \textbf{17}, 2

\bibitem{Damour:2016gwp}
Damour~T 2016 \textit{
Phys. Rev. D} \textbf{94}, no.10, 104015


\bibitem{Buonanno:1998gg}
Buonanno~A and Damour~T 1999 \textit{
Phys. Rev. D} \textbf{59}, 084006

\bibitem{Buonanno:2005xu}
Buonanno~A, Chen~Y and Damour~T 2006 \textit{
Phys. Rev. D} \textbf{74}, 104005


\bibitem{MacKay:2024qxj}
MacKay~N~M 2025 \textit{
Class. Quant. Grav.} \textbf{42}, 24, 245003

\bibitem{MacKay:2025uyg}
MacKay~N~M,
[arXiv:2508.07499 [gr-qc]].


\bibitem{griffiths}
Griffiths~D (\textit{Introduction to Elementary Particle Physics}, Weinheim, John Wiley \& Sons, 1987).

\bibitem{Stewart:1991}
Stewart~J (\textit{Advanced General Relativity}, Cambridge, Cambridge University Press, 1991)
ISBN 0-521-44946-4. 


\bibitem{Feynman:2002}
Feynman~R~P, Morinigo~F~B, Wagner~W~G, Hatfield~B, and Pines~D (\textit{Feynman Lectures on Gravitation}, Westview Press Inc., 2002)
ISBN 978-0-8133-4038-8.

\bibitem{Goldberger:2004jt}
Goldberger~W~D and Rothstein~I~Z 2006 \textit{
Phys. Rev. D} \textbf{73}, 104029 

\bibitem{Goldberger:2006bd}
Goldberger~W~D and Rothstein~I~Z 2006 \textit{
Gen. Rel. Grav.} \textbf{38}, 1537

\bibitem{Goldberger:2007hy}
Goldberger~W~D  2007 
[arXiv:hep-ph/0701129 [hep-ph]].

\bibitem{Kol:2007bc}
Kol~B and Smolkin~M 2008 \textit{
Class. Quant. Grav.} \textbf{25}, 145011 

\bibitem{Goldberger:2009qd}
Goldberger~W~D and Ross~A 2010 \textit{
Phys. Rev. D} \textbf{81}, 124015 

\bibitem{Foffa:2013qca}
Foffa~S and Sturani~R 2014 \textit{
Class. Quant. Grav.} \textbf{31}, no.4, 043001 

\bibitem{Rothstein:2014sra}
Rothstein~I~Z 2014 \textit{
Gen. Rel. Grav.} \textbf{46}, 1726 

\bibitem{Porto:2016pyg}
Porto~R~A 2016 \textit{
Phys. Rept.} \textbf{633}, 1

\bibitem{Levi:2018nxp}
Levi~M 2020 \textit{
Rept. Prog. Phys.} \textbf{83}, no.7, 075901 

\bibitem{Rafie-Zinedine:2018izq}
Rafie-Zinedine~S 
[arXiv:1808.06086 [hep-th]].

\bibitem{Mogull:2020sak}
Mogull~G, Plefka~J and Steinhoff~J 2021 \textit{
JHEP} \textbf{02} 048 

\bibitem{Aoki:2024boe}
Aoki~K, Cristofoli~A and Huang~Y~t 2025 \textit{
JHEP} \textbf{01}, 066

\bibitem{LIGOScientific:2017bnn}
Abbott~B~P, LIGO Scientific and VIRGO,~ \textit{et al.}  2017 \textit{
Phys. Rev. Lett.} \textbf{118}, no.22 221101 
[erratum: 2018 \textit{Phys. Rev. Lett.} \textbf{121}, no.12 129901]


\bibitem{Nakanishi:1979fg}
Nakanishi~N and Ojima~I 1979 \textit{
Phys. Rev. Lett.} \textbf{43}, 91

\bibitem{Cho:2021gvg}
Cho~H~T and Hu~B~L 2002 
\textit{Phys. Rev. D} \textbf{105}, no.8 086004 


\bibitem{Moffat:1996fu}
Moffat~J~W 1997 \textit{
Phys. Rev. D} \textbf{56}, 6264

\bibitem{Hu:1994ep}
Hu~B~L and Matacz~A 1995 \textit{
Phys. Rev. D} \textbf{51}, 1577

\bibitem{DeWitt:1967uc}
DeWitt~B~S 1967 \textit{
Phys. Rev.} \textbf{162} 1239


\bibitem{Blas:2020och}
Blas~D, Martin Camalich~J and Oller~J~A 2022 \textit{
Phys. Lett. B} \textbf{827} 136991 

\bibitem{Delgado:2022uzu}
Delgado~R~L, Dobado~A and Espriu~D 2022 \textit{
EPJ Web Conf.} \textbf{274} 08010 

\bibitem{Herrero-Valea:2022lfd}
Herrero-Valea~M, Koshelev~A~S and Tokareva~A 2022 \textit{
Phys. Rev. D} \textbf{106}, no.10 105002

\bibitem{lang}
van Kampen~N~G (\textit{Stochastic Processes in Physics and Chemistry}, Amsterdam, Elsevier, 1992)

\bibitem{Yuvan2021}
Yuvan~S and Bier~M 2021 \textit{
Phys. Rev. E} \textbf{104} 014119 

\bibitem{Yuvan2022ent}
Yuvan~S and Bier~M 2022 \textit{
Entropy} \textbf{24} 189

\bibitem{Yuvan2022sym}
Yuvan~S, Bellardini~N, and Bier~M 2022 \textit{
Symmetry} \textbf{14} 1042 

\bibitem{Bier2024}
Bier~M 2024 \textit{
Preprints} 2024010282 

\bibitem{MacKay:2024}
MacKay~N~M, 
[arXiv:2406.16117 [cond-mat.stat-mech]]. 


\bibitem{Gundlach:1993tp}
Gundlach~C, Price~R~H and Pullin~J 1994 \textit{
Phys. Rev. D} \textbf{49}, 883


\bibitem{Chow:2004}
Chow~B and Knopf~D (\textit{The Ricci Flow: An Introduction}, Providence, R.I.: American Mathematical Society, 2004)
ISBN 0-8218-3515-7.


\bibitem{Kerr:1963ud}
Kerr~R~P 1963 \textit{
Phys. Rev. Lett.} \textbf{11}, 237

\bibitem{Boyer:1967}
Boyer~R~H and Lindquist~R~W 1967 \textit{
J. Math. Phys.} \textbf{8} (2), 265

\bibitem{Chandrasekhar:1985kt}
Chandrasekhar~S (\textit{The Mathematical Theory of Black Holes}, Oxford Press, 1985)
ISBN: 9780198503705.


\bibitem{dInverno:1992gxs}
d'Inverno~R 1992

\bibitem{Henry:1999rm}
Henry~R~C 2000,
\textit{Astrophys. J.} \textbf{535}, 350 
doi:10.1086/308819
[arXiv:astro-ph/9912320 [astro-ph]].

\bibitem{Visser:2007fj}
Visser~M,
[arXiv:0706.0622 [gr-qc]].

\bibitem{Rovelli:1987df}
Rovelli~C and Smolin~L 1988 \textit{
Phys. Rev. Lett.} \textbf{61} 1155

\bibitem{Rovelli:1989za}
Rovelli~C and Smolin~L 1990 \textit{
Nucl. Phys. B} \textbf{331} 80

\bibitem{Rovelli:1994ge}
Rovelli~C and Smolin~L 1995 \textit{
Nucl. Phys. B} \textbf{442} 593
[erratum: 1995 \textit{Nucl. Phys. B} \textbf{456} 753]

\bibitem{Rovelli:2004}
Rovelli~C (\textit{Quantum Gravity}, Cambridge, Cambridge University Press, 2004)

\bibitem{wolfram}
Wolfram Language 1994. ``FourierCosTransform" \textit{Wolfram Language \& System Documentation Center} ({https://reference.wolfram.com/language/ref/FourierCos Transform.html})

\bibitem{Hubble:1929ig}
Hubble~E 1929 \textit{
Proc. Nat. Acad. Sci.} \textbf{15} 168

\bibitem{Loutrel:2018ssg}
Loutrel~N, Liebersbach~S, Yunes~N and Cornish~N 2019 \textit{
Class. Quant. Grav.} \textbf{36}, no.1 01

\bibitem{Driesse:2024feo}
Driesse~M, Jakobsen~G~U, Klemm~A, Mogull~G, Nega~C, Plefka~J, Sauer~B and Usovitsch~J 2025 \textit{
Nature} \textbf{641}, no.8063, 603


\bibitem{Gross:1973id}
Gross~D~J and Wilczek~F 1973 \textit{
Phys. Rev. Lett.} \textbf{30}, 1343

\bibitem{MacKay:2024sgw}
MacKay~N~M,
[arXiv:2412.20169 [gr-qc]].

\end{thebibliography}
\end{document}